\documentclass[preprint,12pt]{elsarticle}

\usepackage{amsmath,amssymb}
\usepackage[ruled,vlined,english]{algorithm2e}
\usepackage[babel=true]{csquotes}
\usepackage{xcolor}
\usepackage{enumitem}
\usepackage{datetime}
\usepackage{subcaption}
\usepackage{setspace}
\usepackage{lineno,hyperref}
\modulolinenumbers[5]
\usepackage{mathrsfs}
\usepackage{amsbsy}
\usepackage{svg}
\usepackage{amsmath,amsfonts,amsthm,bm} 
\usepackage{multirow}
\usepackage{tabularx}
\usepackage{makecell}
\newcolumntype{C}{>{\centering\arraybackslash}X}   

\begin{document}

\begin{frontmatter}



\title{Augmented state estimation of urban settings using intrusive sequential Data Assimilation}

\author[Poitiers]{L. Villanueva}
\author[Lille]{M. Martínez Valero}
\author[Luxembourg]{A. \v Sarki\' c Glumac}
\author[Lille]{M. Meldi}

\address[Poitiers]{Institut Pprime, CNRS -
ISAE-ENSMA - Universit\'{e} de Poitiers, 11 Bd. Marie et Pierre Curie,
Site du Futuroscope, TSA 41123, 86073 Poitiers Cedex 9, France}

\address[Lille]{Univ. Lille, CNRS, ONERA, Arts et Métiers ParisTech, Centrale Lille, UMR 9014- LMFL- Laboratoire de Mécanique des fluides de Lille - Kampé de Feriet, F-59000 Lille, France}

\address[Luxembourg]{University of Luxembourg, Interdisciplinary Centre for Security, Reliability and Trust (SnT), 6 Avenue de la Fonte, Esch-sur-Alzette, 4364 Luxembourg}

\begin{abstract}
A data-driven investigation of the flow around a high-rise building is performed combining heterogeneous experimental samples and RANS CFD. The coupling is performed using techniques based on the Ensemble Kalman Filter (EnKF), including advanced manipulations such as localization and inflation. The augmented state estimation obtained via EnKF has also been employed to improve the predictive features of the model via an optimization of the five free global model constant of the $\mathcal{K}-\varepsilon$ turbulence model used to close the equations. The optimized values are very far from the classical values prescribed as general recommendations and implemented in codes, but also different from other data-driven analyses reported in the literature. The results obtained with this new optimized parametric description show a global improvement for both the velocity field and the pressure field. In addition, some topological improvement for the flow organization are observed downstream, far from the location of the sensors.  

\end{abstract}




\end{frontmatter}


\section{Introduction}
\label{sec::Intro}

Among the open challenges in the field of fluid mechanics, the accurate prediction and control of the turbulent flows around bluff bodies is a timely subject in the current development of automated, data-informed urban areas. A bluff body is an immersed solid (such as a vehicle, or a building) for which the interaction of the surrounding flow and its shape is responsible for the emergence of large, energetic wakes. When high Reynolds numbers are considered, the global aerodynamic interactions are characterized by complex concurring phenomena such as shear layers, flow separation, and reattachment and recirculation regions. The predictions of such features is a complex task, owing to the large range of active dynamic scales that can be observed in fully developed turbulence. Computational resources required to completely represent turbulent flows via direct numerical simulations are prohibitive for Reynolds numbers observed in realistic applications dealing with urban settings. Reduced-order Computational Fluid Dynamics (CFD) such as RANS \cite{Pope2000_cambridge,Wilcox2006_DCW} can provide a description of complete urban areas with affordable resources, but the accuracy of such prediction is strongly affected by the features of the turbulence model needed to close the dynamic equations. Such models, which are driven by a number of coefficients classically determined via empiric approaches, usually fail in representing interactions of different physical phenomena triggered by turbulence.

Experimental approaches, which rely on measurement that can be obtained by various techniques, such as pressure sensors and hot wires, can provide a virtually exact characterization of the flow features, in the form of pressure and velocity measurements. However, experimental data may be local in space and time, and a full volume representation of flows is prohibitively expensive. In addition, experiments may be affected by difficulties in sensor positioning, which can preclude sampling in sensitive regions of the flow.



Studies in the last two decades have tried to create a solid network between numerical simulation and experiments, in order to exploit the intrinsic advantages of both methods. Among the several proposals in the literature, Data Assimilation \cite{Daley1991_cambridge} is naturally fit to combine experimental and numerical data, in order to obtain a more accurate prediction of the flow. Sequential DA uses tools from probability and statistics to target physical states with minimal uncertainty (state estimation), once different sources of information and their related level of confidence are provided. Among these methods, one can include the Kalman Filter (KF) \cite{Kalman1960_jbe} and its ensemble version, the Ensemble Kalman Filter (EnKF) \cite{Evensen2009_Springer,Asch2016_SIAM} which is de facto among the most powerful tools available for DA. The EnKF can obtain a precise state estimation but also use this physical state to train an underlying model, such as CFD, to better perform in operative conditions.


In this article, RANS simulation is augmented via the integration of experimental data, for the flow around a rectangular building. The augmentation is performed by optimizing a number of global free constants that determine the behaviour of the turbulence closure. Both time-averaged pressure and time-averaged velocity, which are sampled at a limited number of sensors, are used for this purpose. Despite the fact that a number of data-driven analyses to optimize RANS modelling are reported in the literature, using DA \cite{Xiao2016_jcp,Zhang2020_cf} or tools based on machine learning \cite{Duraisamy2019_arfm,Schmelzer2020_ftc}, most of those approaches employ high-fidelity numerical data as reference. The reason why is that numerous additional difficulties are expected when using experimental results, which include overfitting which can lead to the model divergence. It will be shown that approaches based on the EnKF, owing to the smoothing characteristics of the filter, are suitable for robust integration of experimental data within the reduced-order CFD formalism. 

In section \ref{sec::Num}, the numerical strategies and the algorithms used in this work are presented. This includes a description of the numerical solver used as well as a presentation of the data-driven strategies, which are integrated into a specific C++ library. In section \ref{sec:DA_exp}, the setup of the DA analysis is presented. The different techniques which will be compared are detailed. In section \ref{sec:rez} the results obtained are compared with data from a high-fidelity simulation. At last, in section \ref{sec:conclusions} the final remarks are drawn, and future perspectives are discussed.

\section{Numerical Ingredients}
\label{sec::Num}

\subsection{Numerical code: OpenFOAM}
Numerical simulations in this work are performed using a C++ open-source library known as \textit{OpenFOAM}. This library includes a number of solvers based on Finite Volume (FV) discretization \cite{Ferziger1996_springer}, as well as a number of utilities for preprocessing, postprocessing, and data manipulation. Owing to the free license and the very large number of modules available, allowing for extended multi-physics analyses, this code has been extensively used in the literature for research work in fluid mechanics \cite{Constant2017_cf,Meldi2012_pof}. 

For this work, the FV numerical discretization is performed for the RANS Navier-Stokes equations for incompressible flows and Newtonian fluids \cite{Pope2000_cambridge}:

\begin{eqnarray}
\overline{u}_j \frac{\partial \overline{u}_i}{\partial x_j} &=& - \frac{\partial \overline{p}}{\partial x_i} + \frac{\partial \overline{\tau}_{ij}}{\partial x_j} - \frac{\partial \tau^T_{ij}}{\partial x_j} \qquad i=1,2,3 \label{eq:Mom}\\
\nabla^2 \overline{p} &=& - \frac{\partial \overline{u}_{j}}{\partial x_i} \frac{\partial \overline{u}_{j}}{\partial x_i} - \frac{\partial}{\partial x_i} \left( \frac{\partial \tau^T_{ij}}{\partial x_j} \right) \label{eq:Poisson}
\end{eqnarray}

where eq. \ref{eq:Mom} is the momentum equation and eq. \ref{eq:Poisson} is the Poisson equation. The variables used are the velocity $\mathbf{u}=[u_1,u_2,u_3]$, the normalized pressure $p$, the viscous stress tensor $\tau_{ij}$ (which is modelled using the Newtonian fluid hypothesis) and the Reynolds stress tensor $\tau^T_{ij}$. The overbar indicates the average operation performed to obtain eqs. \ref{eq:Mom} - \ref{eq:Poisson}. Within the RANS framework, a turbulence closure must be used for $\tau^T_{ij}$. The $\mathcal{K} - \varepsilon$ model \cite{Launder1974_lhmt,Wilcox2006_DCW} uses the eddy viscosity hypothesis to create a link between $\tau^T_{ij}$ and the gradient of the averaged velocity $\overline{\mathbf{u}}$:
\begin{equation}
   - \tau^T_{ij} = 2 \nu_T \overline{S}_{ij} - \frac{2}{3} \mathcal{K} \delta_{ij}
    \label{eq:ReyStressConst}
\end{equation}
where $\nu_T$ is the turbulent viscosity, $\mathcal{K}$ is the turbulent kinetic energy and $\overline{S}_{ij}$ is the mean strain rate:
\begin{equation}
   \overline{S}_{ij} = \frac{1}{2} \left( \frac{\partial \overline{u}_i}{\partial x_j} + \frac{\partial \overline{u}_j}{\partial x_i} \right) 
    \label{eq:meanStrainRate}
\end{equation}

In the $\mathcal{K} - \varepsilon$ model, $\nu_T$ is expressed as an algebraic function of $\mathcal{K}$ and the energy dissipation rate $\varepsilon$:
\begin{equation}
\nu_T = C_{\mu} \frac{\mathcal{K}^2}{\varepsilon}    
\end{equation}
where $C_{\mu}$ is a model constant to be calibrated. To close the problem, two model equations for $\mathcal{K}$ and $\varepsilon$ must be included:    
\begin{eqnarray}
\frac{\partial \mathcal{K}}{\partial t} + \overline{u}_j \frac{\partial \mathcal{K}}{\partial x_j} &=& \frac{\partial}{\partial x_j} \left( (\nu+\frac{\nu_T}{\sigma_\mathcal{K}}) \frac{\partial \mathcal{K}}{\partial x_j} \right) + \mathcal{P} - \varepsilon \\
\frac{\partial \varepsilon}{\partial t} + \overline{u}_j \frac{\partial \varepsilon}{\partial x_j} &=& \frac{\partial}{\partial x_j} \left( (\nu+\frac{\nu_T}{\sigma_\mathcal{\varepsilon}}) \frac{\partial \varepsilon}{\partial x_j} \right) + C_{\varepsilon 1} \frac{\varepsilon}{\mathcal{K}} \mathcal{P} - C_{\varepsilon 2} \frac{\varepsilon^2}{\mathcal{K}}
\end{eqnarray}
where the production term $\mathcal{P}=\nu_T \overline{S}^2$, $\overline{S}=\sqrt{2 \overline{S}_{ij} \overline{S}_{ij}}$. The model is complete once the five constants $C_\mu$, $C_{\varepsilon 1}$, $C_{\varepsilon 2}$, $\sigma_{\mathcal{K}}$ and $\sigma_{\varepsilon}$ are determined. Launder and Sharma \cite{Launder1974_lhmt} provided values which were determined via the analysis of academic test cases, such as the free decay of homogeneous isotropic turbulence of the turbulent plane channel. However, these coefficients are not constants, but they are a function of the local dynamics of the flow and their interaction with global features of the flow (see discussion in Refs. \cite{Margheri2014_caf,Xiao2019_pas,Duraisamy2019_arfm}). 

\subsection{Observation: experimental data obtained in wind tunnels}\label{S:OD}
\begin{figure}[ht]
	\centering
		\includegraphics[scale=0.65]{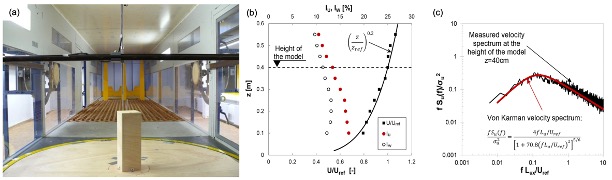}
	\caption{Wind tunnel test section used to produce experimental data}
	\label{FIG:WindTunnelModelAndInflow}
\end{figure}

The experiments are conducted in the atmospheric boundary-layer wind tunnel of the Ruhr-University Bochum, Germany. The wind tunnel has a cross-section of $1.6\times1.8 m$ and a test section length of 9.4 m. Fig. \ref{FIG:WindTunnelModelAndInflow} a) shows the wooden building model mounted on a rotating table in the wind tunnel. 

The boundary layer flow is generated in the wind tunnel using both spires at the tunnel inlet and roughness elements. The mean wind profile matches that of a power law with the exponent of 0.2 as shown in Fig.\ref{FIG:WindTunnelModelAndInflow} b). This is representative of the terrain category II \citep{EN2005} simulating realistic conditions of the flow around isolated high-rise buildings. 

\begin{figure}[ht]
	\centering
		\includegraphics[scale=1.05]{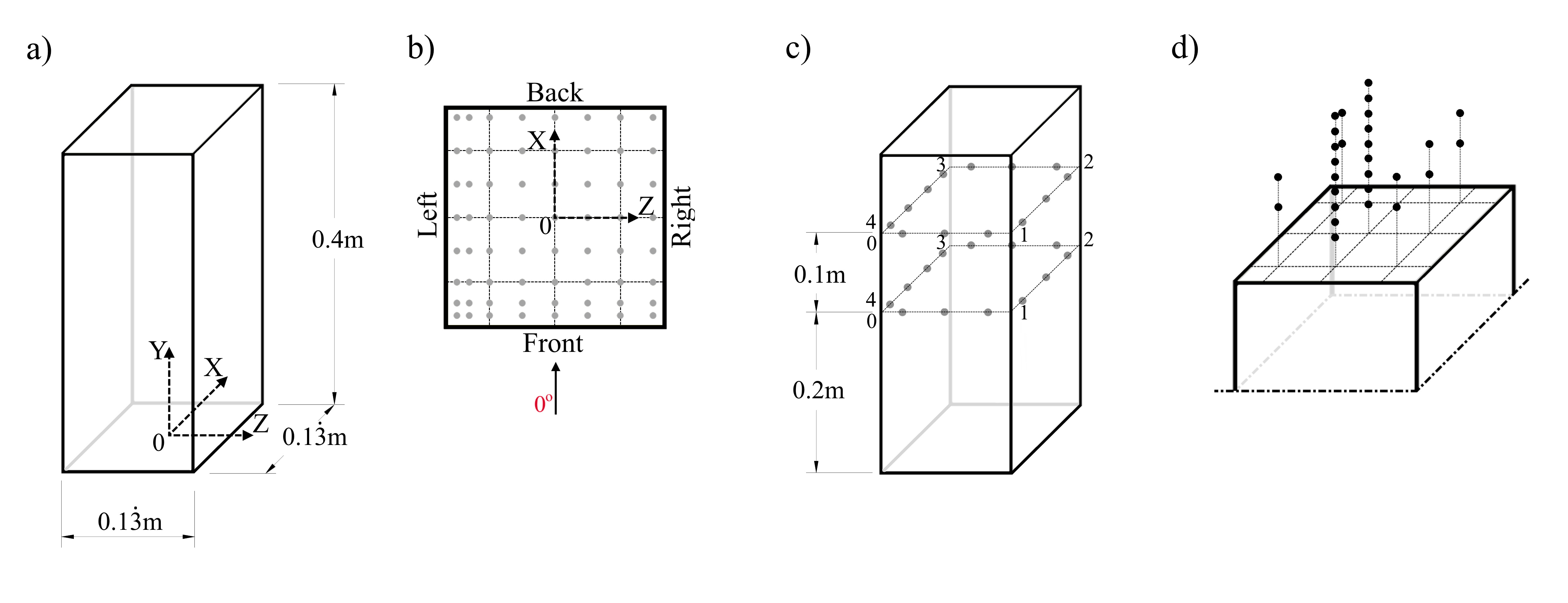}
	\caption{Geometry of the high-rise building with (a) main dimensions and coordinate system; (b) top view with pressure tap locations; (c) facades with pressure tap locations (d) velocity observations measured over the rooftop.}
	\label{FIG:ModelAndMeasurementPoints}
\end{figure}

The sampled data is heterogeneous as different sensors are used to capture features of the velocity field on the roof and pressure measurements on the surface of the building. The velocities above the roof are mainly measured at three different heights ($z/D=0.075$, $0.3$, and $0.45$) above the points marked in red in Fig. \ref{FIG:ModelAndMeasurementPoints} a). In addition, above the centre of the roof, marked P36 in Fig. \ref{FIG:ModelAndMeasurementPoints}, ten heights are considered with the spacing of $z/D=0.075$. The measurements are performed using a hot-wire anemometer, which consists of two cross wires allowing to measure both stream-wise and vertical velocity components. All velocity data are sampled with the frequency of $2000$ Hz. 

In addition to the velocity measurements, the surface pressure is also sampled at different locations. The surface pressure on the roofs of the model are measured using 64 pressure taps, distributed as shown in \ref{FIG:ModelAndMeasurementPoints} b) and c) marked with light gray circles. Surface pressures are acquired with a sampling frequency of 1000 Hz using a multi-channel simultaneous scanning measurement system. The tubing effects are numerically compensated \citep{Neuhaus2010}. More details about the wind tunnel experimentation and the analysis of the flow abound high-rise building, with the special focus on above the roof are presented in \citep{Hemida2020}. In this analysis, experimental data is used to improve the predictive capabilities of stationary RANS models. Therefore, time series available for the velocity components and the pressure have been averaged in time.

\begin{figure}[ht]
	\centering
		\includegraphics[scale=0.6]{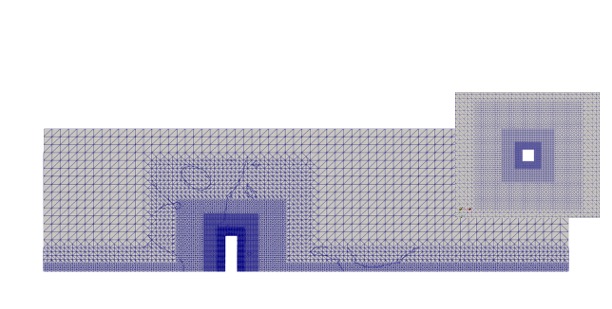}
	\caption{View of the grid used for the RANS calculations. The central vertical plane and an horizontal plane are shown.}
	\label{FIG:Grid}
\end{figure}

\subsection{Test case}
\label{test_case}

The considered high-rise building case is a numerical representation of the wind tunnel tests \citep{Hemida2020}. The model has a square cross-section with edges $B=133.33 mm$, and the height of the building $H=400 mm$, which represents a 120 m tall building in the full-scale. The building has a flat roof, and 0° wind direction is investigated so that the asymtpotic velocity is alligned with the streamwise direction $x$. The lateral direction is $y$ and the normal direction is $z$.

The dimensions of the computational domain are chosen adopting the best practice guidelines given by \cite{Tominaga2008}. The upstream domain length is $5H$. The resulting dimensions of the domain are length ($x$) $\times$ width ($y$) $\times$ height ($z$) $15.5H \times 4.5H \times 4H = 6.2 \times 1.8 \times 1.6m^3$. For the $z$ direction, the height has been chosen to match to the height of the wind tunnel. 

A structured grid is used near the high-rise building surfaces, as shown in Fig. \ref{FIG:Grid}. The distance from the center point of the wall adjacent cell to the building leads to an average $y^+ = 141$ and minimum $y^+ = 40$ which ensures that such point is placed in the logarithmic layer. The total number of mesh elements used to discretize the domain is equal to $513 \, 266$ cells. 

A grid dependency study was performed by comparing the results against a finer grid. The finer grid is composed of 4.3 million cells, characterized by a spatial resolution that is 2 times higher near the building model than the coarse case. The mean pressure predicted by the coarse and fine grid simulations is compared at the locations of the pressure taps on the roof. The comparison showed that 86\% of the points on the building surface have a relative difference below 10\%.

In the simulations the inlet boundary conditions, i.e. mean velocity $\overline{\mathbf{u}}$, the turbulent kinetic energy $\mathcal{K}$ and the turbulence dissipation rate $\varepsilon$, are based on the incident vertical profiles of the mean wind speed $U$ and longitudinal turbulence intensity $I_u$
The turbulent kinetic energy $\mathcal{K}$ is calculated from $U$ and $I_u$: 
\begin{align}
\mathcal{K}(z) &= a(I_u(z)u(z))^2 \label{eq:k}\\
\varepsilon(z) &= \frac{u^{*3}_{ABL}}{\kappa (z+z_0)} \label{eq:eps}
\end{align}
where $a \in[0.5, \, 1.5]$ \cite{Norton2010}. In this study, $a = 1$ is chosen, as recommended by \citep{Tominaga2008}. The turbulence dissipation rate $\epsilon$ is given by Eq.\ref{eq:eps}, with the von Karman constant $\kappa = 0.42$.
 The SIMPLE algorithm was used for pressure-velocity coupling. Classical choices have been performed for the numerical schemes. First-order upwind schemes have been used the convection terms, while second order centered shcemes have been used for viscous terms. Pressure interpolation from cell center to face center has been obtained via second order linear schemes native of the OpenFOAM solver. 

\subsection{Data Assimilation: Ensemble Kalman Filter}
Data assimilation (DA) \cite{Daley1991_cambridge,Asch2016_SIAM} is a family of tools allowing to combine several sources of information to obtain an \textit{augmented prediction} exhibiting increased accuracy. Classical applications usually rely on:
\begin{itemize}
\item a \textit{model}, which provides a (quasi) continuous representation of the physical phenomenon investigated. Physics-based models such as CFD solvers are an example of \textit{model} for fluid mechanics applications 
\item some \textit{observation}, which is usually more accurate of the model, but it is local in space and time. In fluid mechanics, this data may come from high-fidelity numerical simulations or from experiments
\end{itemize}
The \textit{augmented prediction} obtained via manipulation of the sources of information can also be actively used to infer an optimized parametric description of the \textit{model}, with the aim to obtain a predictive tool that can provide accurate predictions without having to rely on \textit{observation}. DA has been traditionally used
in environmental and weather sciences, but applications in fluid mechanics have seen a rapid rise in recent times \cite{Foures2014_jfm, Rochoux2014_nhess, Xiao2016_jcp, Meldi2017_jcp, Meldi2018_ftc, Labahn2019_pci, Chandramouli2020_jcp, Zhang2020_cf, Mons2021_jfm, Moldovan2021_jcp}.
A great variety of methods exists, but two groups can be identified \cite{Asch2016_SIAM, carrassi2018_wcc}:
\begin{description}[topsep=0.4cm] 
    \item[Variational methods:] Methods for which the goal is to minimize a cost function applied for the case studied. This minimum, which is usually reached via parametric optimization of the model, corresponds to an accurate flow state.   
    \item[Statistical (sequential) methods:] Methods for which the goal is to minimize the variance of the solution (i.e. increase the confidence in the prediction). 
\end{description}

Variational methods such as the 4DVar have been extensively used for application in fluid mechanics \cite{Artana2012_jcp,Foures2014_jfm,Mons2017_jfm,Chandramouli2020_jcp} in particular with steady state simulations. While sequential tools are supposedly more appropriate for the prediction of nonstationary phenomena, applications to steady flows are reported in the literature \cite{Meldi2017_jcp,Zhang2020_cf,Zhang2020_jcp}. In the present work, we will focus on tools derived from the Kalman Filter, a well known sequential method. 

\subsubsection{The Kalman Filter}
The Kalman filter (KF) \cite{Kalman1960_jbe} is a sequential DA method based on the Bayes theorem. It provides a solution to the linear filtering of time-dependent discrete data. The classical formulation for the analysis of a physical quantity $\mathbf{x}$ relies on the combination of results produced via a discrete model $\mathbf{M}$, which is linear in the original KF, and some observation $\mathbf{y}$. Within the framework of KF, both the model and the observation are affected by errors/uncertainties, which are here referred to as $v$ and $w$, respectively. One of the central hypotheses of the Kalman Filter is that these uncertainties can be accurately described by a Gaussian distribution i.e. $v=\mathcal{N}(0,\mathbf{Q})$ and $w=\mathcal{N}(0,\mathbf{R})$. $\mathbf{Q}$ and $\mathbf{R}$, which also are a function of time, represent the variance of the model and of the observation, respectively. Considering that these errors can be described by a Gaussian distribution, the solution is completely determined by the first two moments of the state i.e. the physical quantity $\mathbf{x}$ and the error covariance matrix $\mathbf{P}=\mathbb{E}((\mathbf{x}-\mathbb{E}(\mathbf{x}))(\mathbf{x}-\mathbb{E}(\mathbf{x}))^T)$. Let us consider the time advancement of the physical system between the instant $k$ and $k+1$. For the latter, observation $\mathbf{y}_{k+1}$ is available. In this case, the data assimilation procedure consists of two phases (See Algorithm~\ref{alg:KF}):
\begin{enumerate}
    \item A \textit{forecast} step (superscript $f$), where the physical state and the error covariance matrix at the time $k$ are advanced in time using the model:
    $\mathbf{x}_{k+1}^f=\mathbf{M} \mathbf{x}_{k}$
    
    $\mathbf{P}_{k+1}^f=\mathbf{M} \mathbf{P}_{k} \mathbf{M}^T + \mathbf{Q}_{k+1}$
    \item An \textit{analysis} step, where observation and forecast are combined to obtain the \textit{augmented prediction}:
    
    $\mathbf{K}_{k+1}=\mathbf{P}_{k+1}^f \mathbf{H}^T \left(\mathbf{H} \mathbf{P}_{k+1}^f \mathbf{H}^T + \mathbf{R}_{k+1}\right)^{-1}$
    
    $\mathbf{x}_{k+1}^a=\mathbf{x}_{k+1}^f + \mathbf{K}_{k+1} ( \mathbf{y}_{k+1} - \mathbf{H} \mathbf{x}_{k+1})$
    
    $\mathbf{P}_{k+1}^a=(\mathbf{I} - \mathbf{K}_{k+1} \mathbf{H}) \mathbf{P}_{k+1}^f$
\end{enumerate}

where $\mathbf{H}$ is a linear operator mapping the model results to the observation space and $\mathbf{K}$ is the Kalman gain. This matrix takes into account the correlations between the values of the state vector and the values of the observations, and it is the central element providing the final state estimation of the physical system.  

The main drawbacks of this algorithm for complex applications in fluid mechanics are that i) it is designed for linear models $\mathbf{M}$ and ii) the size of $P$ is directly linked with the number of degrees of freedom of the problem investigated. While the first issue can be bypassed with ad-hoc improvements of the data-driven strategy, which are included for example in the \textit{extended} KF \cite{Asch2016_SIAM, carrassi2018_wcc}, the second one is more serious. In fact, $P$ must be advanced in time like the physical variables. In addition, during the \textit{analysis} phase, extended manipulation of $\mathbf{P}$ is required, including a matrix inversion. For the number of degrees of freedom used in CFD, which are usually in the range $10^6 - 10^8$, this leads to prohibitive requirements in terms of RAM and computational resources.    

\begin{algorithm}
\setstretch{1.2}
\caption{Algorithm for The Kalman Filter}
\label{alg:KF}
\textbf{Forecast steps}\\
\nl \qquad$\mathbf{x}_{k+1}^f = \mathbf{M}\mathbf{x}_k^a$\\
\nl \qquad$\mathbf{P}_{k+1}^f = \mathbf{M}\mathbf{P}_k^a\mathbf{M}^T + \mathbf{Q}_{k+1}$\\
\textbf{Analysis steps}\\
\nl \qquad$\mathbf{K}_{k+1} = \mathbf{P}_{k+1}^f\mathbf{H}^T\left(\mathbf{H}\mathbf{P}_{k+1}^f\mathbf{H}^T + \mathbf{R}_{k+1}\right)^{-1}$\\
\nl \qquad$\mathbf{x}_{k+1}^a = \mathbf{x}_{k+1}^f + \mathbf{K}_{k+1}(\mathbf{y}_{k+1}-\mathbf{s}_{k+1}^f)$\\
\nl \qquad$\mathbf{P}_{k+1}^a = (\mathbf{I}-\mathbf{K}_{k+1}\mathbf{H})\mathbf{P}_{k+1}^f$
\end{algorithm}

\subsubsection{The (stochastic) Ensemble Kalman Filter} 
The Ensemble Kalman filter (EnKF) \cite{Evensen2009_IEEE,Asch2016_SIAM} is a popular data-driven strategy based on the KF which provides an efficient solution to the issues previously discussed. The idea is that the error covariance matrix $\mathbf{P}$ is not advanced in time anymore, but it is approximated via an ensemble of model runs. This strategy allows to fully account for non linearity of the model and it virtually eliminates the computational burdens associated with the manipulation of $\mathbf{P}$. The complete structure of the EnKF, which is summarized in the Algorithm \ref{alg:EnKF}, is now discussed.

The EnKF relies on $N_e$ realizations of the model, which is the model ensemble. At a given instant $k$, the realizations can be assembled in a state matrix $\mathbf{X}_S$ of size $[N,N_e]$, where $N$ is the number of degrees of freedom of the physical problem investigated. Therefore, keeping the usage of the superscripts $f$ and $a$ introduced for the KF, each column of $\mathbf{X}_S$ corresponds to the state $\mathbf{x}_{i,k+1}^f$ of the $i^{th}$ member, where $i \in [1, N_e]$.    An approximation of the error covariance matrix $\mathbf{P}_e$ can be obtained exploiting the hypothesis of statistically independence of the ensemble members:
\begin{equation}
    \mathbf{P}_e^f = \mathbf{X}^f(\mathbf{X}^f)^T
\end{equation}
where $\mathbf{X}^f$ is the anomaly matrix which represents the deviation of all the values of the state vectors from 
their ensemble mean:
\begin{equation}
    \mathbf{X}_{k+1} = \frac{\mathbf{x}_{i,k+1}-\overline{\mathbf{x}_{k+1}}}{\sqrt{N_e-1}} \; , \qquad \overline{\mathbf{x}_{k+1}} = \frac{1}{N_e}\sum_{i = 1}^{N_e}\mathbf{x}_{i,k+1}
\end{equation}

The sampled observation, which consist of $N_o$ elements, is also expanded to obtain $N_e$ sets of values. To do so, a Gaussian noise based on the covariance matrix of the measurement error $\mathbf{R}_{k+1}$ is added to the observation vector:
\begin{equation}
    \label{eqn:EnKF_obs}
    \mathbf{y}_{i,k+1} = \mathbf{y}_{k+1} + \mathbf{e}_{i,k+1},\; with \; \mathbf{e}_{i,k+1} \thicksim \mathcal{N}(0, \mathbf{R}_{k+1})
\end{equation}
Finally, the model realizations are projected to the observation space $N_e$ times, similarly to the classical KF:
\begin{equation}
    \label{eqn:EnKF_Projobs}
    \mathbf{s}_{i,k+1} = \mathbf{H} \mathbf{x}_{i,k+1}
\end{equation}

All of these elements together allow for the determination of the Kalman gain:
\begin{eqnarray}
\mathbf{S}_{k+1}= \frac{\mathbf{s}_{i,k+1}-\overline{\mathbf{s}_{k+1}}}{\sqrt{N_e-1}} \; , \qquad \overline{\mathbf{s}_{k+1}} = \frac{1}{N_e}\sum_{i = 1}^{N_e}\mathbf{s}_{i,k+1} \\
\mathbf{E}_{k+1}= \frac{\mathbf{e}_{i,k+1}-\overline{\mathbf{e}_{k+1}}}{\sqrt{N_e-1}} \; , \qquad \overline{\mathbf{e}_{k+1}} = \frac{1}{N_e}\sum_{i = 1}^{N_e}\mathbf{e}_{i,k+1} \\
    \label{eqn:EnKF_gain_R}
    \mathbf{K}_{k+1} = \mathbf{X}_{k+1}^f(\mathbf{S}_{k+1}^f)^T \left[\mathbf{S}_{k+1}^f(\mathbf{S}_{k+1}^f)^T + \mathbf{E}_{k+1}(\mathbf{E}_{k+1})^T\right]^{-1}
\end{eqnarray}
In an infinite ensemble size $\mathbf{E}_{k+1}(\mathbf{E}_{k+1})^T$ tends to the matrix $\mathbf{R}$ of the Kalman filter. In practice the size is limited, thus the product of the perturbations is simplified by the diagonal matrix $\mathbf{R}_{k+1}$ gaining simplification and computational cost \cite{carrassi2018_wcc, Hoteit2015_mwr}. 
In addition, $\mathbf{P}_e$ can be directly estimated from the ensemble members for each analysis phase, and there is no need for memory storage/time advancement. 

Finally, the update of the state vectors is then performed in the same way as in the classical KF, the only difference being that $N_e$ updates have to be performed: 
\begin{equation}
    \mathbf{x}_{i,k+1}^a = \mathbf{x}_{i,k+1}^f + \mathbf{K}_{k+1}(\mathbf{y}_{i,k+1}-\mathbf{s}_{i,k+1}^f)
\end{equation}


The EnKF can also be used to optimize the parametric description of the model. The underlying idea is that the parameters are updated at the end of the analysis phase so that the model can provide a more accurate prediction of the physical phenomenon investigated, reducing the difference between the model predicted state and the final state estimation. Several strategies are proposed in the literature \cite{Asch2016_SIAM} and, among those, one showing efficiency for a relatively small set of parameters (referred to as $\theta$) and easy to implement is the so-called \textit{extended state}. In this strategy, the steps of the EnKF are performed for a state $\mathbf{x}^\star$ which is defined as:
\begin{equation}
\mathbf{x}^\star = \begin{bmatrix} 
	\mathbf{x} \\
	\theta
	\end{bmatrix}
\end{equation}

That is, the state used for the EnKF includes both the physical state and the parametric description of the model. For this very simple algorithm, the size of the global state is now equal to $N^{\star}=N + N_\theta$, where $N_\theta$ is the number of parameters to be optimized. This modification brings a negligible increase in computational costs if $N_\theta << N$, and it allows us to obtain simultaneously an updated state estimation and optimized parametric description for the model at the end of the analysis phase.

\begin{algorithm}[H]
    \caption{Algorithm for the Ensemble Kalman Filter}
    \label{alg:EnKF}
    \textbf{For $N_e$ the number of member in the ensemble, $i = 1,...,N_e$:}
    \nl Advancement in time of the state vectors:\\
    \qquad $\boldsymbol{x}_{i,k+1}^f = \mathcal{M}\boldsymbol{x}_{i,k}^a$ \\
    \nl Creation of an observation matrix from the observation data by introducing errors:\\
    \qquad$\boldsymbol{y}_{i,k+1} = \boldsymbol{y}_{k+1} + \boldsymbol{e}_{i}$, with $\boldsymbol{e}_{i} \thicksim \mathcal{N}(0,\boldsymbol{R})$\\
    \nl Calculation of the predicted observation:\\
    \qquad$\boldsymbol{s}_{i,k+1}^f = \mathcal{H}\boldsymbol{x}_{i,k+1}^f$\\
    \nl Calculation of the ensemble means:\\
    \qquad$\overline{\boldsymbol{x}_{k+1}^f} = \frac{1}{N_e}\sum_{i = 1}^{N_e}\boldsymbol{x}_{i,k+1}^f$,\,
    $\overline{\boldsymbol{s}_{k+1}^f} = \frac{1}{N_e}\sum_{i = 1}^{N_e}\boldsymbol{s}_{i,k+1}^f$,\\
    \qquad$\overline{\boldsymbol{e}_{k+1}} = \frac{1}{N_e}\sum_{i = 1}^{N_e}\boldsymbol{e}_{i,k+1}$\\
    \nl Calculation of the anomaly matrices:\\
    \qquad$\boldsymbol{X}_{k+1} = \frac{\boldsymbol{x}_{i,k+1}-\overline{\boldsymbol{x}_{k+1}}}{\sqrt{m-1}}$,\,
    $\boldsymbol{S}_{k+1} = \frac{\boldsymbol{s}_{i,k+1}-\overline{\boldsymbol{s}_{k+1}}}{\sqrt{m-1}}$,\,
    $\boldsymbol{E}_{k+1} = \frac{\boldsymbol{e}_{i,k+1}-\overline{\boldsymbol{e}_{k+1}}}{\sqrt{m-1}}$\\
    \nl Calculation of the Kalman gain:\\
    \qquad$\boldsymbol{K}_{k+1} = \boldsymbol{X}_{k+1}^f(\boldsymbol{S}_{k+1}^f)^T \left[\boldsymbol{S}_{k+1}^f(\boldsymbol{S}_{k+1}^f)^T + \boldsymbol{R}_{k+1}\right]^{-1}$\\
    \nl Update of the state matrix:\\
    \qquad$\boldsymbol{x}_{i,k+1}^a = \boldsymbol{x}_{i,k+1}^f + \boldsymbol{K}_{k+1}(\boldsymbol{y}_{i,k+1}- \boldsymbol{s}_{i,k+1}^f)$
\end{algorithm}

\subsubsection{Inflation}
The classical EnKF exhibits a number of shortcomings such as sampling errors due to the limited amount of members available in the ensemble. This is especially true for applications in fluid mechanics and in particular with CFD, where every simulation may need important computational resources and storage space. Therefore, the number of total ensemble members realistically acceptable for three-dimensional runs is around $N_e \in [40, 100]$. As this error is carried over the assimilation steps, one way of reducing this problem is to inflate the error covariance matrix $\mathbf{P}_{k+1}$ by a factor 
$\lambda^2$ \cite{Asch2016_SIAM}. 

This coefficient $\lambda > 1$ drives the so called \emph{multiplicative inflation}, which can be applied to the analyzed state matrix. It is responsible for an increased variability of the state estimation:
\begin{equation}
    \mathbf{x}_{i}^a \longrightarrow \overline{\mathbf{x}^a} + \lambda(\mathbf{x}_{i}^a-\overline{\mathbf{x}^a})
\end{equation}

Clearly, for $\lambda=1$ the results from the classical EnKF are obtained. 

Similarly, the optimization via EnKF of the set of inferred parameters $\theta$ can collapse very rapidly towards a local optimum, providing a sub-optimal result. Inflation can be used to mitigate an overly fast collapse of the parametric description of the model, artificially increasing the variability of the parameters and allowing it to target a global optimum solution.

\subsubsection{Localization}

The classical EnKF establishes a correlation between observation and the degrees of freedom of the model, but this correlation is not affected by the distance between them. In a limited ensemble size like the ones used in CFD, this can lead to spurious effects on the update of the state matrix for large domains. In practice, errors due to the finite ensemble approximations can be significantly larger than the real physical correlation, which naturally decays with distance in continuous systems. Due to the computational limitations to using more members in the ensemble, one way to avoid these spurious effects is to use a corrective multiplicative term to the values of the covariance matrix $\mathbf{P}_{k+1}$ that takes into account the physical distance between observation sensors and mesh elements of the state. This strategy is known as \emph{covariance localization}. Just as the inflation, the localization is effective to improve the accuracy of the calculation and to reduce the probability of divergence of the EnKF.
The principle of covariance localization uses a coefficient-wise multiplication of the covariance matrix $\mathbf{P}_{k+1}$ and a corrective matrix that is here called $\mathbf{L}$.
This type of operation is known as a Schur product, thus it is also called \emph{Schur localization}. This leads to the expression of the localized Kalman gain 
Equation \ref{localized_EnKF}.

\begin{equation}
    \label{localized_EnKF}
    [\mathbf{P}_{k+1}]_{i,j}[\mathbf{L}]_{i,j} \longrightarrow {\mathbf{K}_{k+1}^{loc} = [\mathbf{L}]_{i,j}[\mathbf{X}_{k+1}^f(\mathbf{S}_{k+1}^f)^T]_{i,j}\left([\mathbf{L}]_{i,j}[\mathbf{S}_{k+1}^f(\mathbf{S}_{k+1}^f)^T]_{i,j} + \mathbf{R}_{k+1}\right)^{-1}}
\end{equation}

As the matrix $\mathbf{R}_{k+1}$ has a limited impact on the operation, this expression is simplified for convenience in the algorithm. The localized Kalman gain becomes:
\begin{equation}
    [\mathbf{P}_{k+1}]_{i,j}[\mathbf{L}]_{i,j} \longrightarrow {\mathbf{K}_{k+1}^{loc} = [\mathbf{L}]_{i,j}[\mathbf{K}_{k+1}]_{i,j}}
\end{equation}

The structure of the matrix $\mathbf{L}$ must be set by the user, and it should represent the real physical correlation. In continuous systems, the correlation between physical variables decreases fast in space. Therefore, a generally used structure for the localization matrix is an exponential decay form: 
\begin{equation}
    \label{loc_matrix}
    \mathbf{L}(i,j) = e^{-\Delta^2_{i,j}/\mu}
\end{equation}
where $\Delta_{i,j}$ is the distance between the given observation sensor and the point of evaluation of the model (center of the mesh element in CFD). $\mu$ is a decay coefficient that can be tuned accordingly to the characteristics of the test case.

\subsection{CONES: Coupling OpenFOAM with Numerical EnvironmentS}
CONES (Coupling OpenFOAM with Numerical EnvironmentS) is a C++ library designed to couple the CFD software OpenFOAM with any other kind of open-source code. It is currently employed to carry out sequential DA techniques and, more specifically, advanced data-driven methods based on the EnKF. The communications between the EnKF-based code and OpenFOAM are performed by CWIPI (Coupling With Interpolation Parallel Interface) \cite{Reflox2011_aerlab}, which is an open-source code coupler for massively parallel multi-physics/multi-components applications and dynamic algorithms. 

The main favorable features of CONES in performing DA with OpenFOAM are the following:
\begin{itemize}
    \item It is not needed to modify the installation of OpenFOAM but only compile user-made functions.
    \item The intrusive coupling between the different codes is performed preserving the original structure of the existing CFD solvers. Every CONES related function is contained in a Pstream (Part of OpenFOAM) modified library, hence, data exchange is done at the end of the solver loop by calling specific functions, and the calculation loop remains unmodified.
    \item It is very efficient to exchange information about the physical state and the mesh (arrays of millions of elements can be sent and received simultaneously and rapidly). 
    \item Direct HPC communications between multiple processors, which handle partitions of the numerical simulations and the DA processes.
    \item Simulations and DA are run simultaneously \textit{online} i.e. there is no need to stop the simulations at each analysis phase. This last point allows for enormous gains in terms of computational resources required.
\end{itemize}

The coupler CWIPI developed by ONERA and CERFACS has been chosen due to its powerful management of fully parallel data exchanges based on distributed mesh definition and its ability to interpolate between non-coincident meshes (very useful for some advanced tools based on the EnKF, like the MGEnKF \cite{Moldovan2021_jcp}). Most of its uses are related to gas turbine designs \cite{Duchaine2015_csd, Legrenzi2016_AIAA}, but it has also recently been employed in the field of aeroacoustics with OpenFOAM \cite{Moratilla-Vega2022_cpc}.

\begin{figure}[ht]
    \centering
    \subfloat[{\footnotesize CWIPI between two codes}]{\label{fig:CWIPIaction}{\includegraphics[width=0.45\textwidth]{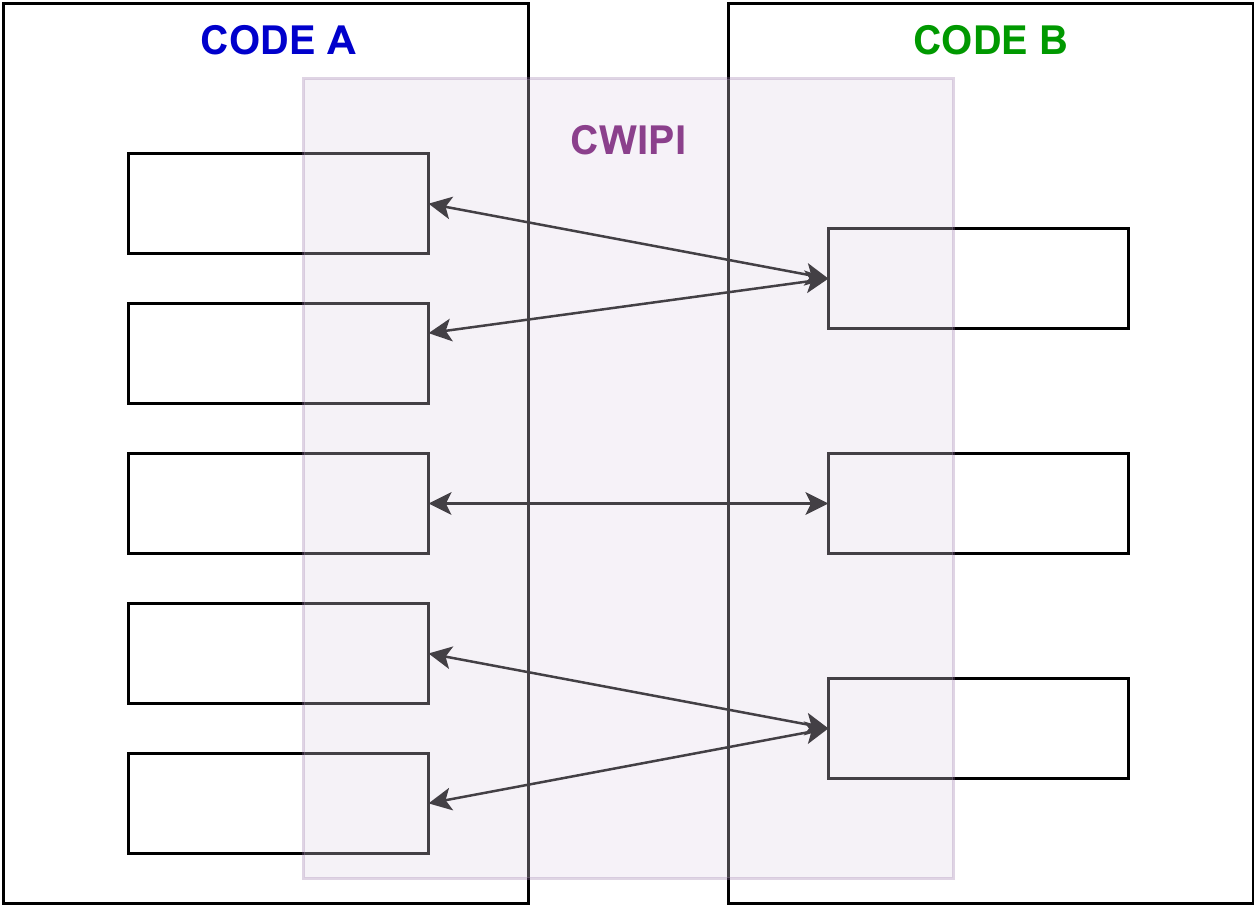}}}\hfill
    \hspace{0.3cm}
    \subfloat[{\footnotesize Main primitives in CWIPI}]{\label{fig:primitives}{\includegraphics[width=0.50\textwidth]{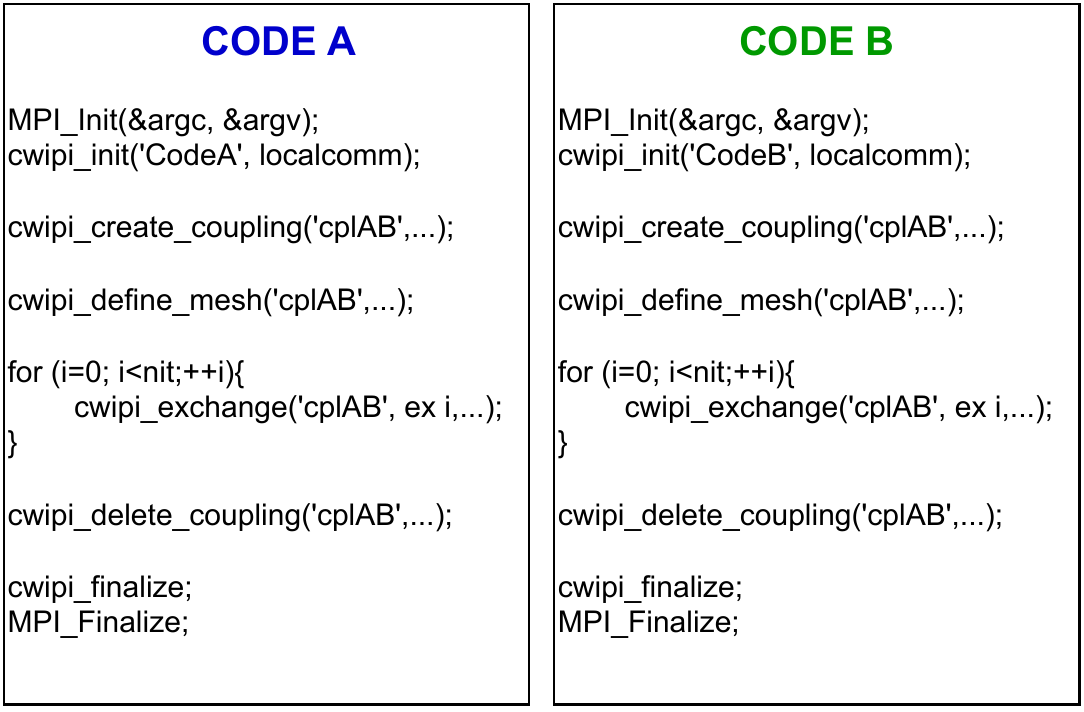}}}
    \caption
    {Functioning of CWIPI}
    \label{fig:CWIPI}
\end{figure}

CWIPI communication protocol is based on the MPI library. Thus, MPI and CWIPI environments must be initialized within the codes. This will allow using the primitives of CWIPI to exchange information between two codes. In Figure \ref{fig:CWIPI} direct communications between two codes through CWIPI and some of the main primitives are illustrated.

\begin{figure}
    \centering
    \includegraphics[width=1\textwidth]{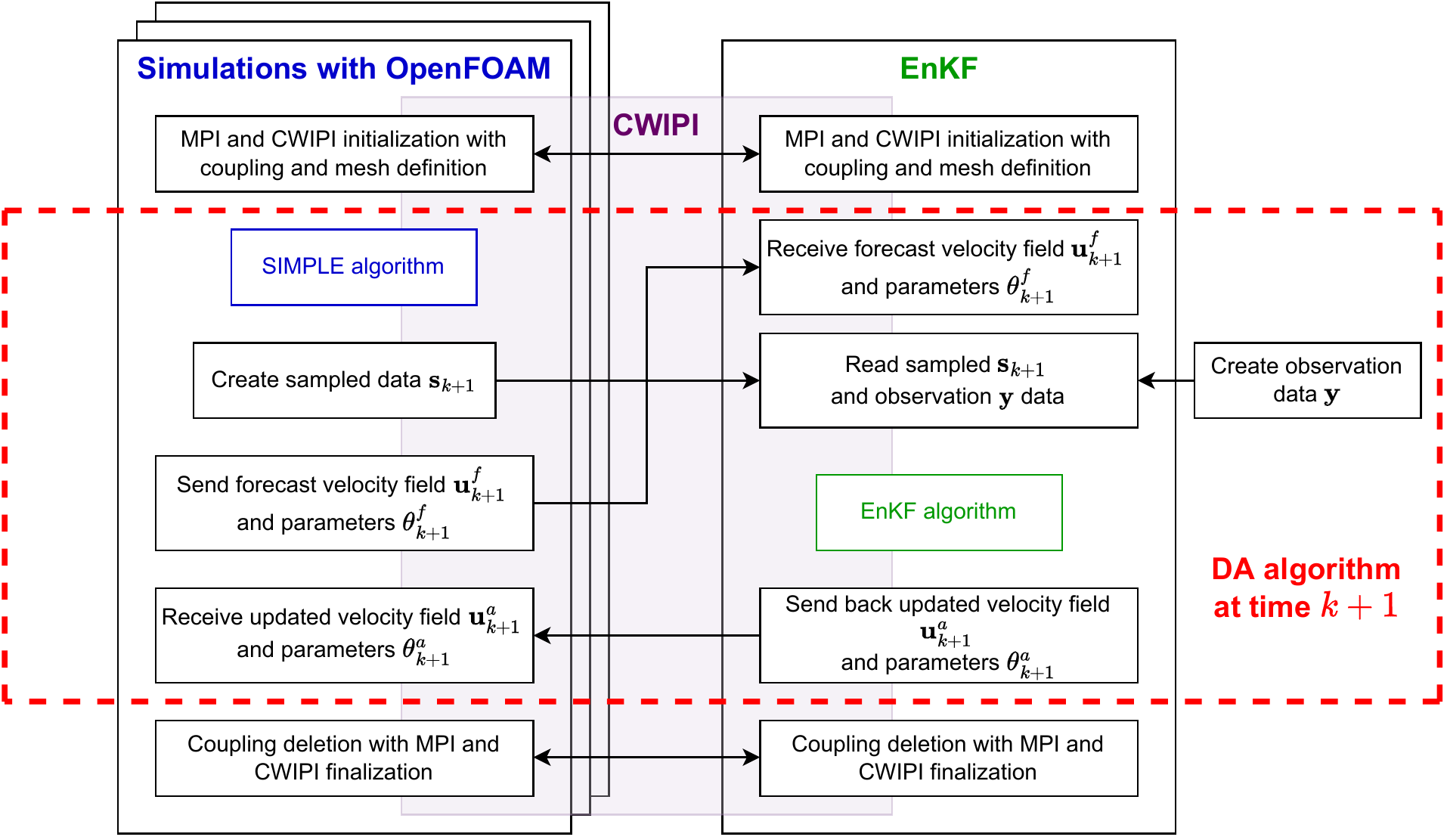}
    \caption{Scheme of CONES for steady simulations}
    \label{fig:CONES}
\end{figure}

In this work, CONES couples the solver SimpleFOAM, which is designed to simulate flows using RANS, with a sequential DA library developed by the team. The structure of a single run is exemplified in Fig. \ref{fig:CONES}. The MPI communications and the coupler CWIPI are initialized in both codes (in OpenFOAM and in the DA library).

Despite the fact that applications of EnKF-based tools are tied with the time advancement of the solution, the application to stationary flows is straightforward. An analysis virtual time window is fixed in terms of the number of iterative steps of the code. Once that number of time steps is performed simultaneously by the $N_e$ ensemble members (CFD runs), they send their information to the EnKF code and wait online for the updated flow field / parametric description. Currently, the information exchanged is the velocity field $\mathbf{u}^f$ and the parameters of the model $\theta^f$. Hence, the state matrix, composed of as many state vectors as members (CFD simulations) in the ensemble, is the one expressed in eq. \ref{eqn:state_vector} for the DA cycle at time $k+1$.

\begin{equation}
\mathbf{x}_{i,k+1} = \begin{bmatrix} \mathbf{u}_{i,k+1} \\ \theta_{i,k+1}
    \end{bmatrix}
\label{eqn:state_vector}
\end{equation}

A piece of additional information provided by the simulations is the set of values $\mathbf{s}_{i,k+1}$, i.e. the projection of the model solution on the coordinates of the sensors for each ensemble member. Considering that the sensor placement does not necessarily comply with the center of a mesh element, interpolation of the flow field has to be performed. OpenFOAM possesses several functions to transform cell-center quantities into particular points. The accuracy of these interpolation methods \cite{Leonard2021_ihpbc} has been taken into account. 

The nature of the observations $\mathbf{y}$ is analyzed in more detail in Section \ref{sec:DA_exp}, but CONES can work with sensors measuring both the pressure $p$ and the velocity field $\mathbf{u}$. In this specific case dealing with stationary simulations, the observation is constant, and it is loaded once, but it could be integrated at each analysis phase in case of analysis of an unstationary flow. Thus, the DA code receives information from the model and the sensors, and it produces an updated set of states and parameters ($\mathbf{u}^a$, $\theta^a$), which are sent back to the OpenFOAM simulations. The pressure $p$ is updated for each ensemble member via a Poisson equation and this complete set of data is used to start a new set of iterative steps. Once the convergence of the model parameters complies with a threshold set by the user from the model is achieved, the coupling is deleted, and both MPI and CWIPI environments are finalized.

\section{DA experiments}
\label{sec:DA_exp}
CONES is here used to study the flow around a building using the numerical test case presented in Sec. \ref{test_case}. In particular, the DA tools are used to optimize the value of the five global constants driving the $\mathcal{K} - \varepsilon$ model, with the aim to minimize the discrepancy between the RANS results and the high-fidelity observation provided. A similar analysis was recently performed by Zhao et al. \cite{Zhao2022_be}, but they used numerical results from a high-fidelity simulation. Here, the observation is taken from time-averaged measures from experiments. 

The first key aspect to take into account is determining a suitable prior state for the velocity and pressure field, as well as for the parametric description. For the latter, values found by Margheri et al. \cite{Margheri2014_cf} using uncertainty propagation of epistemic uncertainties are preferred to the classical values obtained by Launder and Sharma \cite{Launder1974_lhmt}. These baseline values, which are shown in Tab. \ref{tab:Prior_EnKF}, are the initial mean of the $N_e$ ensemble simulations. Each value of the parameters for each CFD run is initially determined using a bounded Gaussian distribution $\mathcal{N}(\mu_N, \sigma_N)$, where $\mu_N$ is the parameter mean value and $\sigma_N$ is chosen to provide a sufficiently large initial variability of the parametric space based on Margheri et al. work \cite{Margheri2014_cf}. The normal distributions are bounded between $\sigma_N$ and $7\sigma_N$, the limit has been empirically set depending on the sensitivity of the coefficients. For example, $C_{\varepsilon1}$ is bounded by $1.25\sigma_N$ but $\sigma_\varepsilon$ is bounded by $7\sigma_N$. The initial physical state for each ensemble is obtained from a single run using the values of the model constants in Ref. \cite{Margheri2014_cf}.  

The number of ensemble members $N_e = 40$ is chosen considering other works in the literature relying on CFD for the model part of the EnKF \cite{katzfuss2016_as, Mons2021_prf, Moldovan2021_jcp}. 


The observation is obtained from time-averaged data from a total of $120$ sensors. Among these, $90$ sensors are pressure taps, and $30$ sensors are hot wires measuring two components of the velocity field, the streamwise velocity $u_x$ and spanwise velocity $u_z$. This adds up to $150$ time-averaged observation values. The data is loaded at the beginning of the analysis phase in the following format: $\mathbf{y}= \begin{bmatrix} u_{x1} & \dots & u_{x30} & u_{z} & \dots & u_{z30} & p_{31} & \dots & p_{120} \end{bmatrix}^T$, and does not change throughout the calculation. Similarly the covariance matrix $R_{k+1}$ is taken constant leading to $R = \sigma_m I$ with $\sigma_m$ the confidence given to the measurement.

Three independent DA experiments are performed. The variations do not deal with details of the model or the observation, but they consider different features of the DA procedure. More precisely, the cases analyzed are: 
\begin{itemize}
    \item Case A: classical EnKF.
    \item Case B: EnKF with covariance localization.
    \item Case C: EnKF with covariance localization and inflation.
\end{itemize}

\begin{table}[!ht]
    \centering
    \begin{tabularx}{\textwidth}{|C|C|C|C|C|}
        \hline
        \multirow{3}{*}{\textbf{Parameter}} &  \multirow{2}{*}{\thead{$\mathbf{\mathcal{K}}$\textbf{-}$\mathbf{\varepsilon}$ \textbf{model}\\ \textbf{default}\\ \textbf{values}}}  & \multicolumn{3}{c|}{\textbf{Prior of the EnKF}}\\ \cline{3-5}
         & & \thead{$\mathbf{\mu_N}$} & \thead{$\mathbf{\sigma_N}$ \\ for cases A,B} & \thead{$\mathbf{\sigma_N}$ \\ for case C}\\ 
         \hline \hline
         $C_\mu$ [-] & 0.09 & 0.1 & 0.01 & 0.005\\
         $C_{\varepsilon1}$ [-] & 1.44 & 1.575 & 0.1 & 0.05\\
         $C_{\varepsilon2}$ [-] & 1.92 & 1.9 & 0.1 & 0.05\\
         $\sigma_\mathcal{K}$ [-] & 1.0 & 1.0 & 0.1 & 0.05\\
         $\sigma_\varepsilon$ [-] & 1.3 & 1.6 & 0.1 & 0.05\\ \hline
    \end{tabularx}
    \caption{Comparison between conventional constants from RANS $\mathcal{K}-\varepsilon$ model and the initial parameters employed for the EnKF ($N_e = 40$)}
    \label{tab:Prior_EnKF}
\end{table}

For case study A and B, the width of the virtual assimilation window has been fixed to $150$ iterations. For C it has been fixed to $100$ considering additional evolution of the parameters caused by the inflation. Those values have been chosen observing results from preliminary analyses, which pointed out how at least $50$ iterations were needed to obtain a clear signature of the new parametric setting over the physical quantities.

\begin{figure}[h!]
    \centering
    \includegraphics[width=0.7\textwidth]{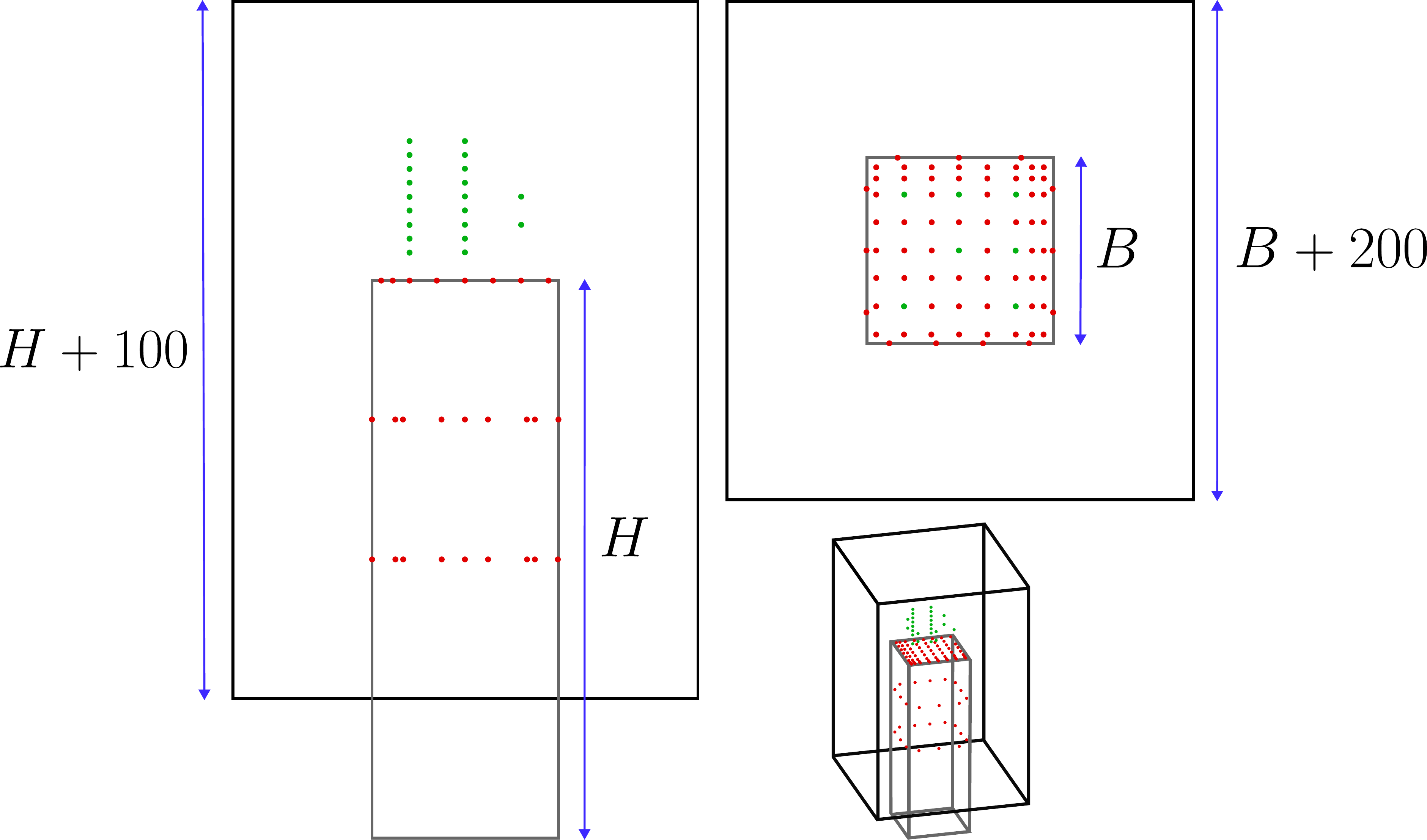}
    \caption{Clipping box used for localization: pressure sensors are represented in red and velocity sensors are displayed in green (data in $mm$) }
    \label{fig:sensors_clipping}
\end{figure}

For localization in cases B and C, the domain is also clipped in a volume sufficiently large around the observations, similar to the way Moldovan et al. \cite{Moldovan2022_jfm} did it for the BARC geometry. Similarly to the physical clipping, that reduces the number of degress of freedom in the EnKF and thus the computational cost, the goal of the localization clipping is also to ensure eliminating regions which are far from the sensors and that would only be marginally updated by the DA procedure. Fig. \ref{fig:sensors_clipping} shows the cutting region defined for the physical localization. The coefficient $\mu$ of the localization matrix $\mathbf{L}$ is set so that the values of the matrix are close to $0$ at the border of the clipping. This avoids discontinuity problems in the physical field. The parameters of the model $\theta$ are not affected by localization. 




\subsection{Case A: classical EnKF}
\begin{figure}[h!]
    \centering
    \includegraphics[width=1\textwidth]{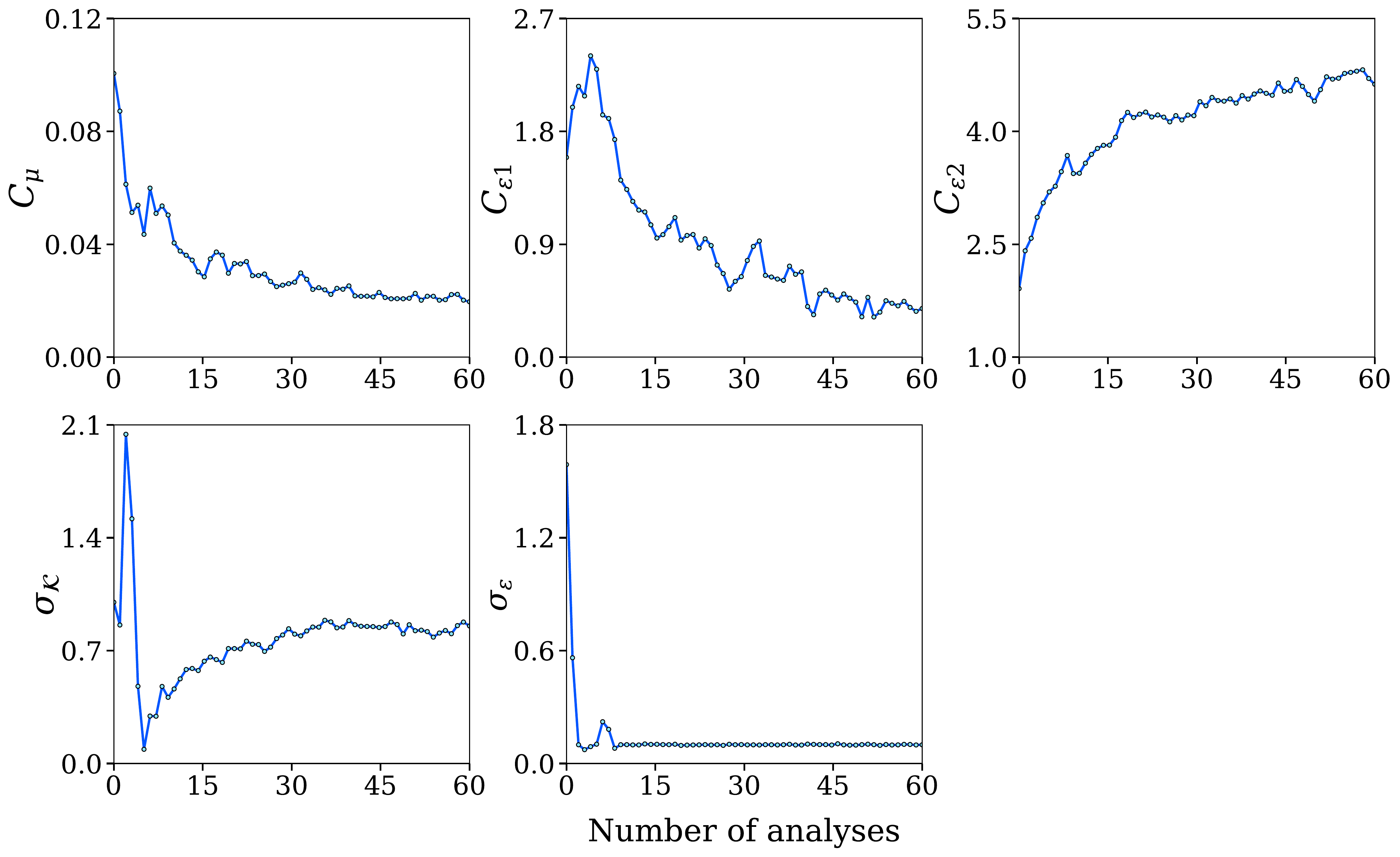}
    \caption{$\mathcal{K} - \varepsilon$ coefficients convergence without inflation and localization}
    \label{fig:coeff_withoutBoth}
\end{figure}

In this first case, the classical Ensemble Kalman Filter is used. The run ends when convergence of the parameters is reached, which is in this case after $60$ analysis phases (i.e. a total of $6000$ CFD iterations). The evolution of the mean value of the five parameters of the $\mathcal{K} - \varepsilon$ model is shown in Fig. \ref{fig:coeff_withoutBoth}. One can see that the final results obtained by the EnKF are significantly different than the baseline values, and that the speed at which the parameter converge is significantly different. In particular, the evolution of $\sigma_{\varepsilon}$ deserves some comments. This coefficient controls the magnitude of the turbulent diffusion term in the equation for $\varepsilon$, $D_{\varepsilon} = \nu_t/\sigma_{\varepsilon}+\nu$, which is associated with non-homogeneous conditions (see Sec. \ref{sec::Num}). The optimization performed by the EnKF targets very low values for $\sigma_{\varepsilon}$ during the calculation, increasing the relevance of $D_{\varepsilon}$ in the equation. However, noise propagated by the Kalman gain can turn the value for some ensemble members to be negative, resulting in a divergence of the calculation. Therefore, we have imposed a constraint for this parameter so that values cannot be lower than a small but positive value prescribed. For the other parameters, one can see that $C_\mu$ and $C_{\varepsilon1}$ converge to a value close to $1/3$ of the initial estimate, $\sigma_\mathcal{K}$ does not exhibit large variations and $C_{\varepsilon2}$ is three times larger. 

\subsection{Case B: EnKF with covariance localization}
\begin{figure}[h!]
    \centering
    \includegraphics[width=1\textwidth]{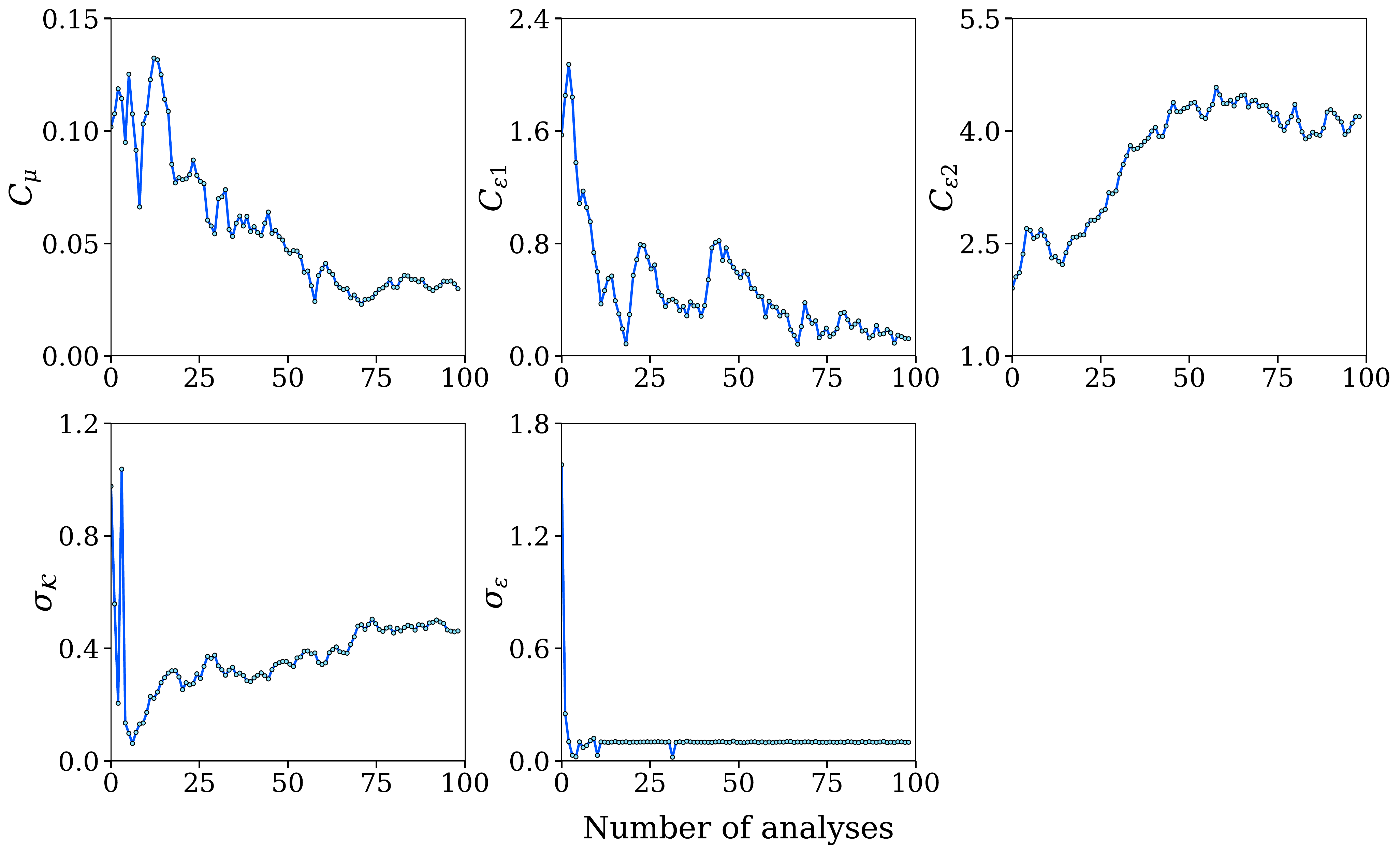}
    \caption{$\mathcal{K} - \varepsilon$ coefficients convergence without inflation but with localization}
    \label{fig:coeff_withoutInfl}
\end{figure}

In this case, the calculation is performed with covariance localization. The evolution of the five coefficients is shown in Figure \ref{fig:coeff_withoutInfl}. The trend and in particular the evolution of $\sigma_\varepsilon$ are similar to the ones observed for Case A. given remarks for $\sigma_{\varepsilon}$ in case A stay the same here. However, larger fluctuations can be observed before convergence. Also, a significantly larger number of iterations ($100$ analysis phases) is required to get a good convergence of the parameters.

\subsection{Case C:  EnKF with both inflation and localization}
\begin{figure}[h!]
    \centering
    \includegraphics[width=1\textwidth]{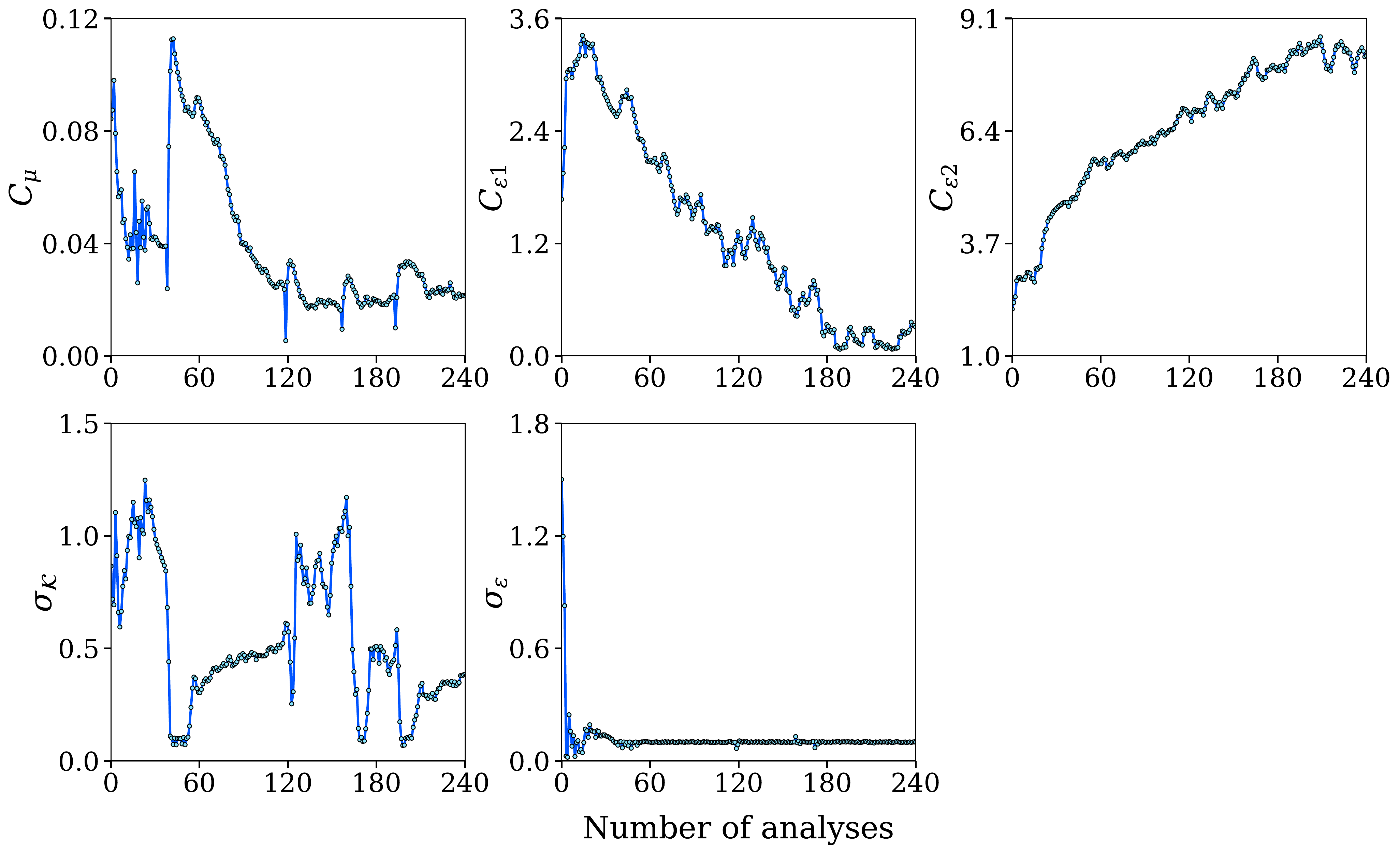}
    \caption{$\mathcal{K}-\varepsilon$ coefficients convergence with inflation and localization}
    \label{fig:coeff_withBoth}
\end{figure}

The DA calculation is here performed relying on deterministic inflation for the model's parameters and covariance localization. This is the most advanced run in terms of complexity of the DA algorithm. The evolution of the five parameters is shown in Figure \ref{fig:coeff_withBoth}. To ensure the robustness of the simulation during the first time steps, the inflation quantifier $\lambda$ is gradually increased from 1.05 to 1.3 and, later, removed to obtain the convergence ($\lambda=1.05$ for $k\in [1,40]$, $\lambda=1.1$ for $k\in [41,120]$, $\lambda=1.2$ for $k\in[121,160]$, $\lambda=1.3$ for $k\in[161,200]$, and $\lambda=1$ for $k>200$). Some coefficients such as $C_\mu$ and $\sigma_\mathcal{K}$ show a higher sensitivity to changes of the value of $\lambda$, highlighting the importance of inflation in identifying a suitable large parametric space for the optimization. For this reason, convergence is reached significantly later in this case. Also, the threshold value for $\sigma_\varepsilon$ is increased here, in order to avoid stability problems that could be easily triggered by the higher variability associated with the parametric inflation.

The impact of the physical prediction of the three different parametric descriptions, which are reported in Table \ref{tab:opti_values}, are investigated in Section \ref{sec:rez}. 

\begin{table}[h!]
    \centering
    \begin{tabularx}{\textwidth}{|C|C|C|C|}
        \hline
        \multirow{2}{*}{\textbf{Parameter}} & \multicolumn{3}{c|}{\textbf{Optimized values}}\\ 
        \cline{2-4}
         & \thead{\textbf{Case A}} & \thead{\textbf{Case B}} & \thead{\textbf{Case C}}\\
         \hline
         $C_\mu$ [-] & 0.021 & 0.032 & 0.020 \\
         $C_{\varepsilon 1}$ [-] & 0.418 & 0.165 & 0.162 \\
         $C_{\varepsilon 2}$ [-] & 4.629 & 4.080 & 7.978 \\
         $\sigma_\mathcal{K}$ [-] & 0.837 & 0.476 & 0.355 \\
         $\sigma_\varepsilon$ [-] & 0.1 & 0.1 & 0.1\\ \hline
    \end{tabularx}
    \caption{$\mathcal{K}-\varepsilon$ model coefficients final values obtained with the EnKF}
    \label{tab:opti_values}
\end{table}

\section{Results}
\label{sec:rez}

Results obtained by the three DA runs are now compared with the prior (classical RANS using $\mathcal{K}-\varepsilon$ model) and the time-averaged experimental results.   

\subsection{Velocity field}
The analysis of the velocity field is performed first. Velocity is an explicit variable in segregated solvers for incompressible flows. Therefore, the performance of the DA strategies can be assessed by the qualitative improvement obtained for the prediction of this quantity.

\begin{figure}[ht]
	\centering
		\includegraphics[scale=1.6]{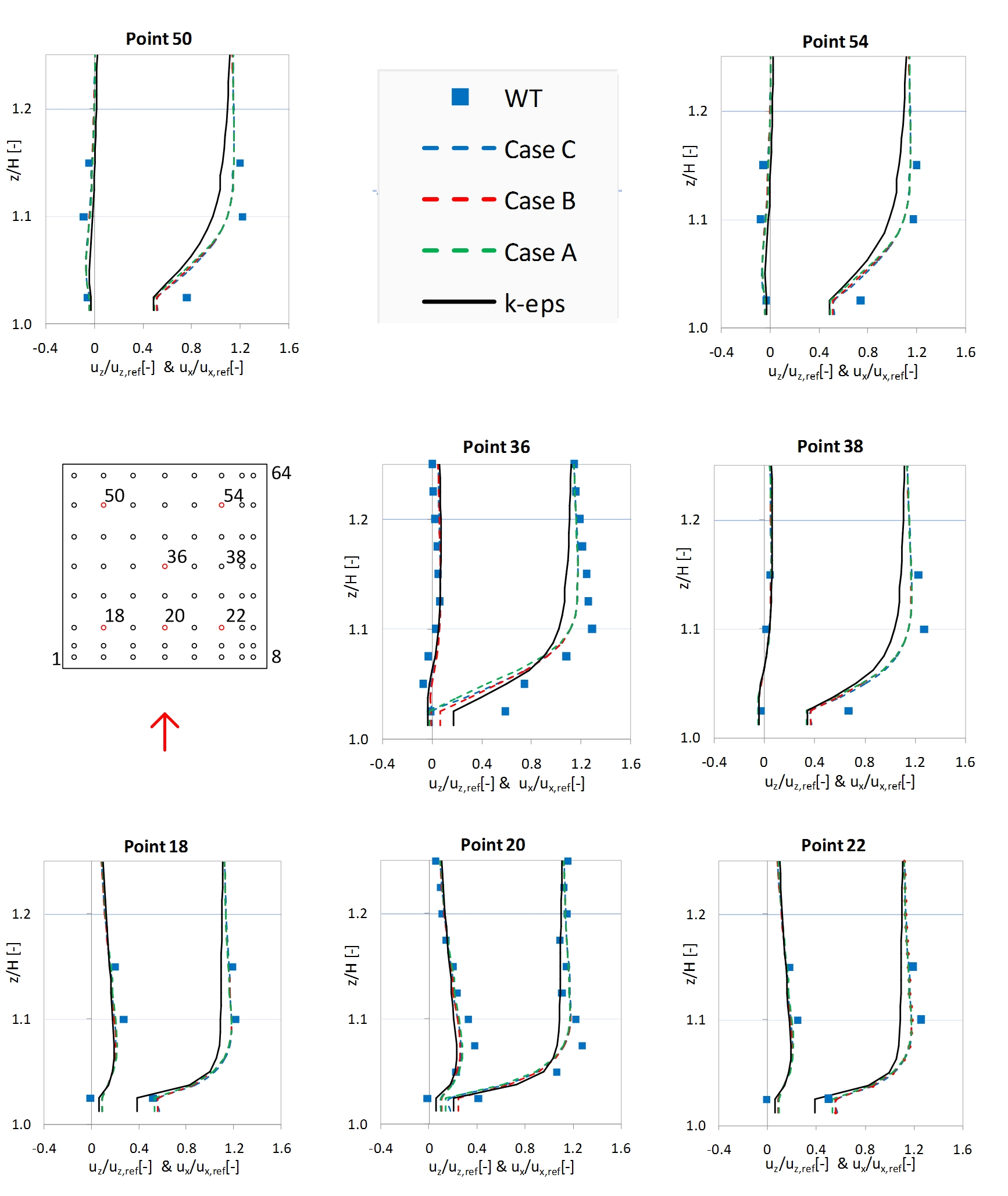}
	\caption{Vertical and streamwise velocity profiles above the marked red locations on the roof (Points: 18, 20, 22, 36, 38, 50, 54); comparison between wind tunnel data (WT), k-epsilon model (k-eps), and three ensembles Kalman filter cases.}
	\label{FIG:VelProfiles}
\end{figure}

Figure \ref{FIG:VelProfiles} shows the streamwise velocity profiles $u_x$ and the normal velocity profiles $u_z$
for several locations corresponding to the positions of the hot wire. For $u_x$, one can see that the accuracy of the predicted field via DA is sensibly improved for each location. Very minor differences can be observed when comparing the three different DA strategies. On the other hand, the prediction of the normal velocity $u_z$ is very similar to the prior. This is not surprising, considering that the match between prior and experimental data is good. An interesting result can be observed for Point $20$, where the maximum difference between the prior and the experimental data is observed for $u_z$. In this case, one can see that the DA prediction is getting closer to the experiments, confirming that the EnKF is able to provide a statistically more accurate prediction of the flow, down to the confidence indicated for the different sources of information 

\begin{figure}[ht]
	\centering
		\includegraphics[scale=1.1]{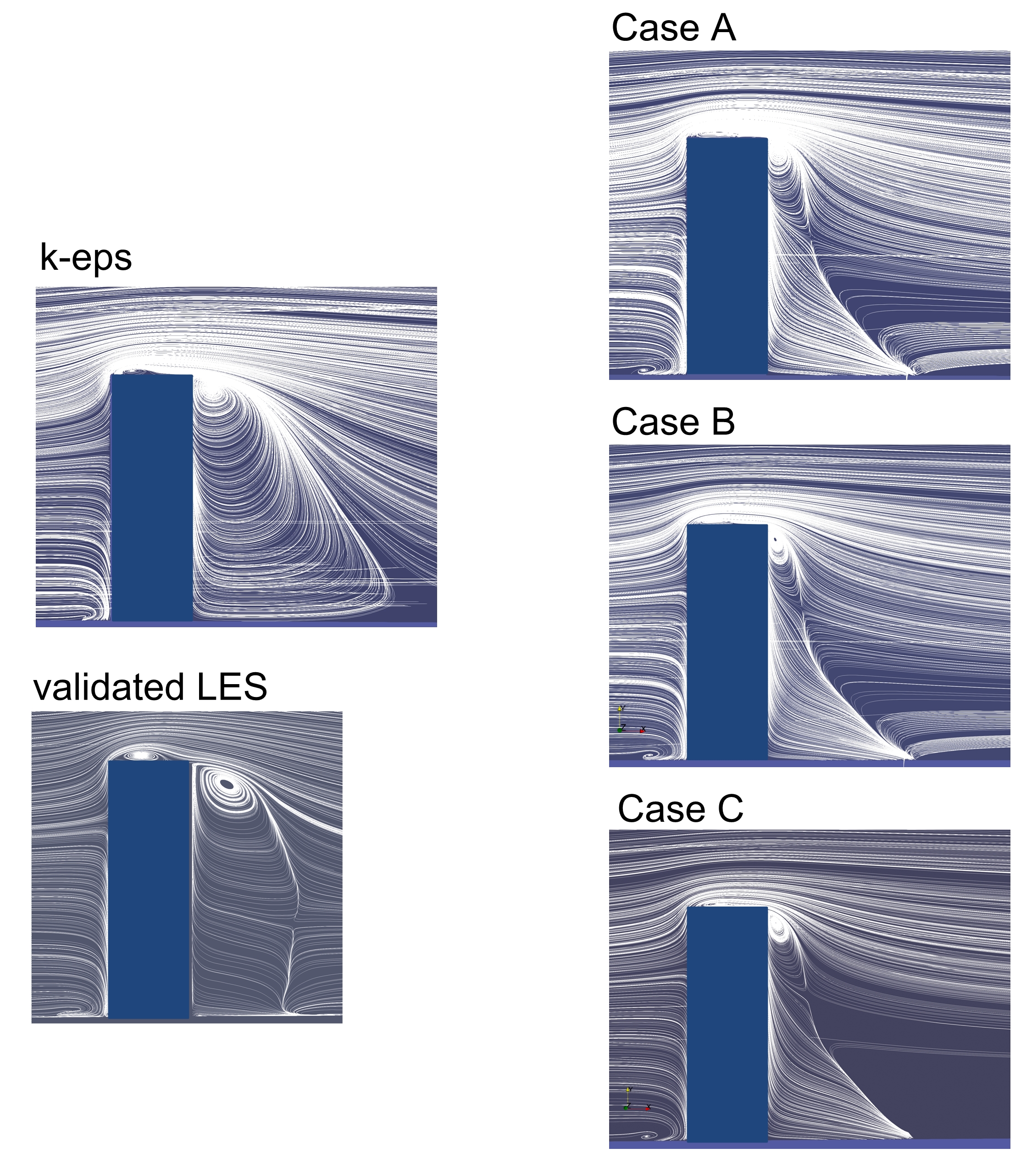}
	\caption{Flow structures at the middle plane; comparison between validated Large Eddy Simulation (validated LES) published in \citep{vranesevic_furthering_2022} and k-epsilon model (k-eps) on the left side of the Figure, and three ensembles Kalman filter cases ot the right side of the Figure.}
	\label{FIG:Streamlines}
\end{figure}

\begin{figure}[ht]
	\centering
		\includegraphics[scale=1]{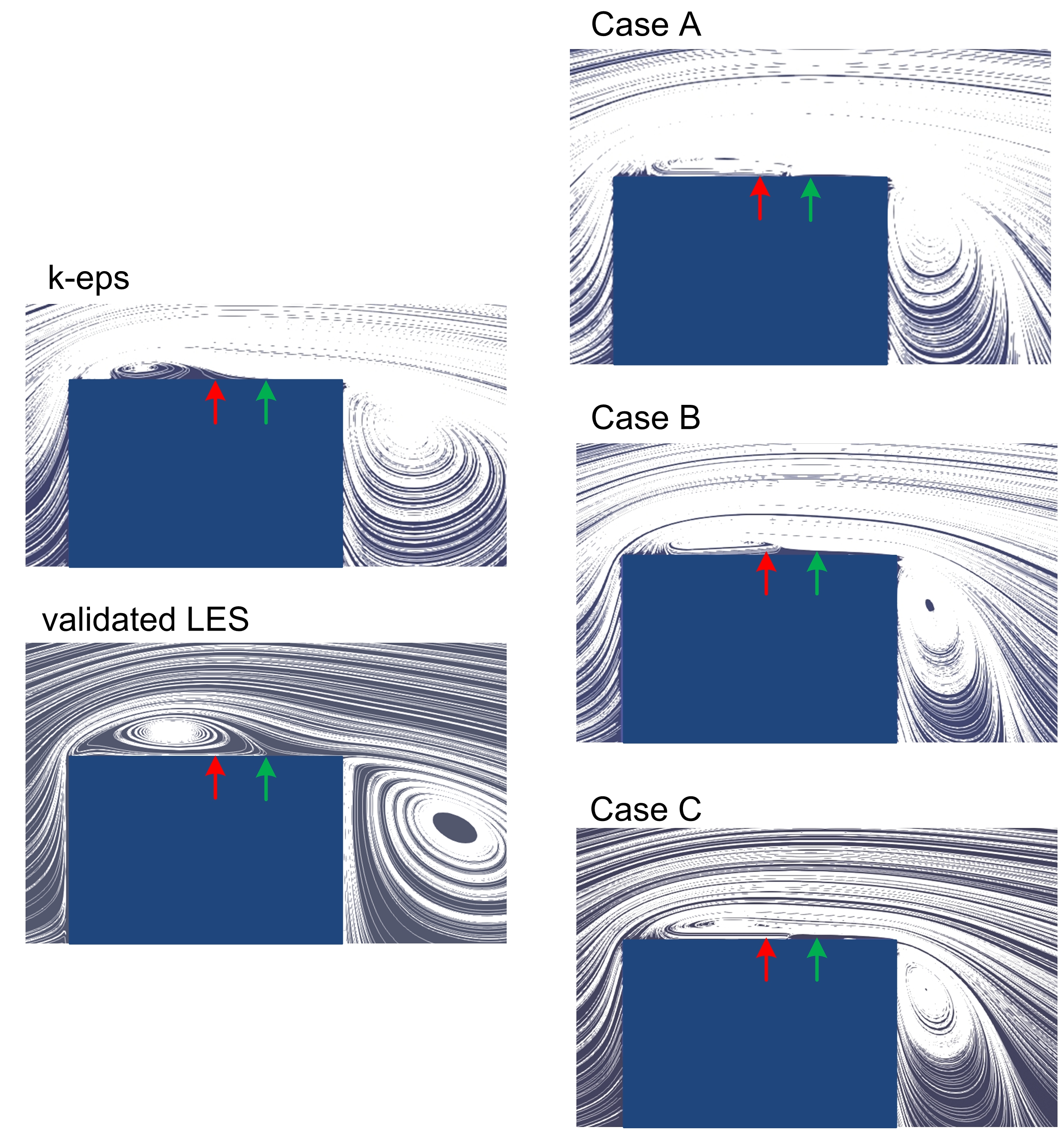}
	\caption{Close view of the flow structures above the roof top, red arrow - indicates the reattachment position of the separation bubble of the simulation with k-epsilon model, green arrow - indicates the reattachment position of the separation bubble of the validated LES simulation}
 \label{FIG:recircBubble}
 \end{figure}

The features of the velocity field are further assessed in figure \ref{FIG:Streamlines}, where streamlines on a vertical plane at the center of the high-rise building are shown. Here, the prior and the DA runs are compared with a validated LES study. One can see that the RANS parametric variation obtained integrating local experimental information in the room region is responsible for a significant reduction of the recirculation region behind the building. While the size of this region is now smaller than the reference LES, similar topological organization can be observed at mid-height, which is not captured by the RANS prior. A zoom of the roof area, which is shown in figure \ref{FIG:recircBubble}, shows how the assimilation of the velocity field is beneficial in improving the prediction of the recirculation region, which is sensibly improved.

\subsection{Pressure field}
\begin{figure}[ht]
	\centering
		\includegraphics[scale=0.7]{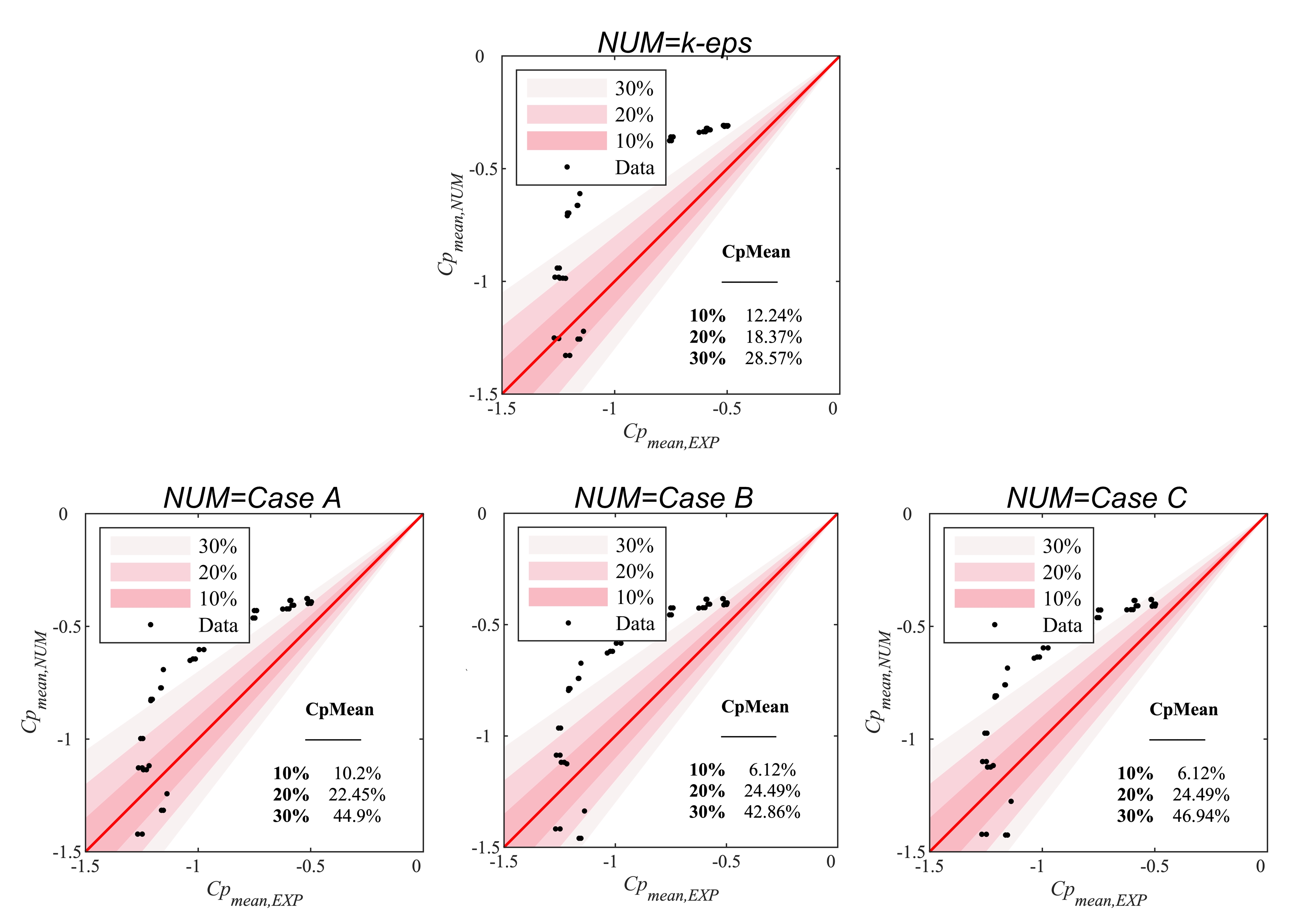}
	\caption{Scatter plot of mean pressure coefficient for all numerical cases with performance metrics-comparison with experimental data.}
	\label{FIG:PMatrix}
\end{figure}

\begin{figure}[ht]
	\centering
		\includegraphics[scale=1.25]{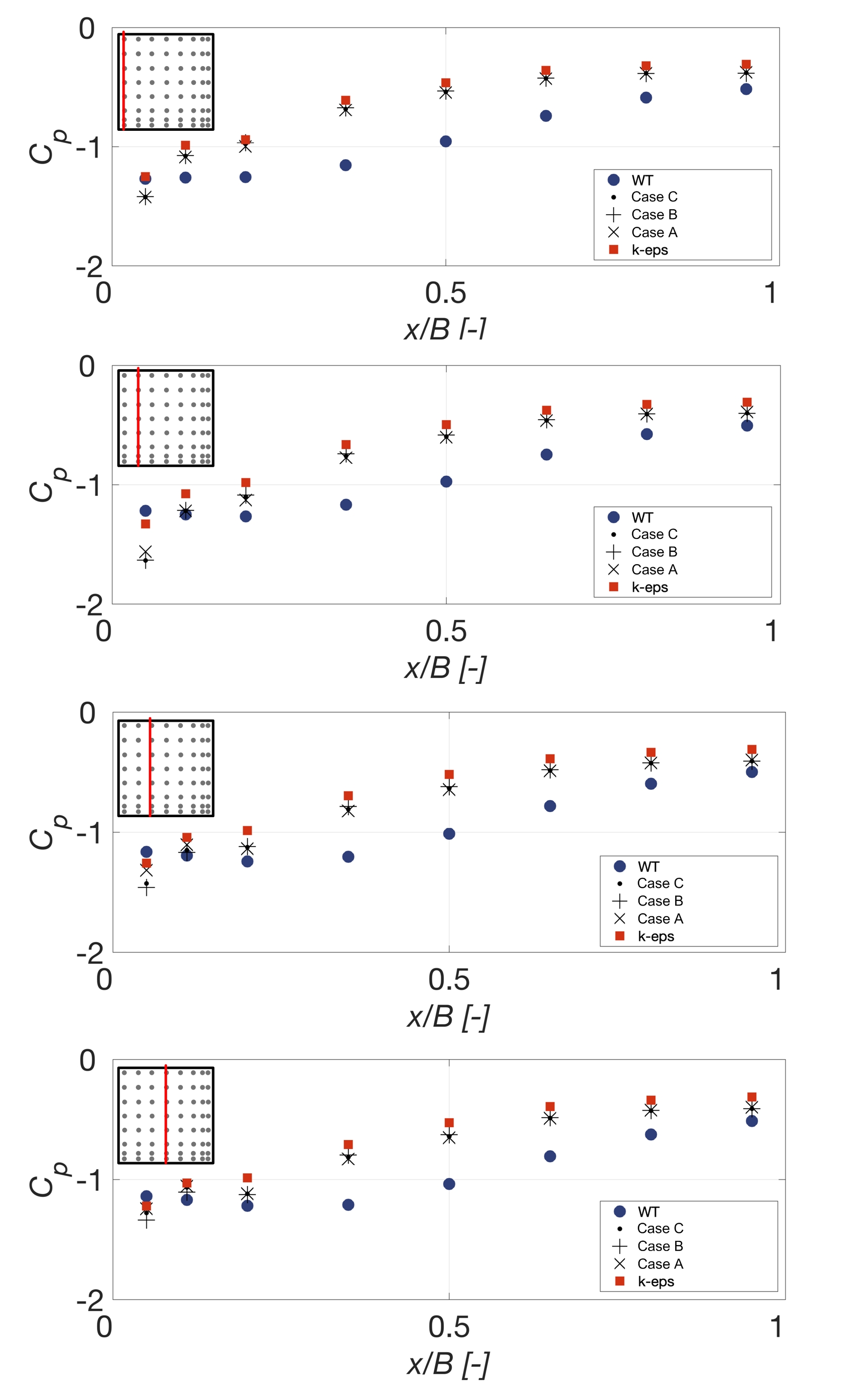}
	\caption{Mean pressure coefficient at pressure tap locations along red lines marked at the roof; comparison between wind tunnel data (WT), k-epsilon model (k-eps), and three ensembles Kalman filter cases}
	\label{FIG:POnLines}
\end{figure}

The behaviour of the pressure field is now investigated. This physical quantity is significantly more difficult to be predicted, because the Poisson equation resolved in the numerical solver uses the pressure as a Lagrangian marker. Therefore, the pressure field is simultaneously used as a physical variable and a Lagrangian constraint to grant incompressibility of the flow. Therefore, the analysis of this quantity is crucial to assess the stability and the precision of the algorithms. In figure \ref{FIG:PMatrix} the mean pressure coefficient is shown in terms of performance metrics comparison with experimental data. If one just has a look at the best performance region for the error $>10\%$, one could be erroneously lead to this that the prior ($12\%$ of the occurrences for less than $10\%$ error) behaves better than the DA runs (between $6\%$ and $10\%$ of the occurrences for less than $10\%$ error). This information is misleading, though, as a large number of occurrences for the DA runs are just outside this interval. In fact, as large margins of error are considered, one can see that the DA runs outperform the prior RANS. For a $20\%$ error threshold, an improvement of around $5\%$ occurrences ($22.5-24.5\%$ against $18.4\%$) is observed with the use of DA. The gap rises to around $15\%$ when a $30\%$ error threshold is considered. This trend is confirmed analyzing the mean pressure coefficients calculated at pressure taps on the roof, shown in figure \ref{FIG:POnLines}. One can see that the DA prediction almost always performs better than the prior, even if gains for the pressure field are less important than what observed for the velocity field.

Some remarks should be spent about the performance of the three DA runs. Despite the differences in the techniques used and the apparently different results obtained for the parameter optimization, the prediction of the physical variables is pretty similar. The values of the model parameters have probably converged towards a robust optimum, where the sensitivity of the solution to further parametric variation is very low. This aspect, which needs further investigation, may indicate that robust optimization can be obtained setting a suitable confidence interval for the observation. In this scenario, the application of localization has proven effective. The reduction of degrees of freedom in the DA process, which significantly decreases the computational resources required for each analysis phase, is not responsible for the degradation of the results. On the other hand, probably because of the features of the parametric optimum region found, the inflation techniques have not improved the results.

\section{Conclusions}
\label{sec:conclusions}

The newly developed platform CONES has been used to perform a data-driven investigation of the flow around a high-rise building. More precisely heterogeneous experimental samples, in the form of data from pressure taps and hot wires, have been integrated with RANS CFD runs, performed using the open-source code OpenFOAM. The coupling has been performed using techniques based on the Ensemble Kalman Filter (EnKF), including advanced manipulations such as localization and inflation. The augmented state estimation obtained via EnKF has also been employed to improve the predictive features of the model via an optimization of the five free global model constant of the $\mathcal{K}-\varepsilon$ turbulence model used to close the equations. Therefore, a relatively small uncertainty space has been chosen, the same employed by Ben Ali et al. \cite{BenAli2022_jweia} using variational data assimilation.

The results have shown that a global improvement has been observed for the physical quantities of investigation, and that results obtained with the different DA strategies are equivalent. For this last point, physical and covariance localization appear to be effective for the study of complex flows. The reduction of degrees of freedom of the DA problem has not affected the quality of the results, while globally reducing the time needed for the data-driven procedures. On the other hand, the usage of inflation has not produced better results, in particular due to the increase of computational resources required.

The analysis of the velocity field shows that the EnKF allows to significantly reduce the error in the streamwise and normal direction of more than $50\%$. The effects of the parametric inference are observed also on the recirculation region behind the building. In this case, the physical topology of the flow becomes more similar to the reference LES validated with experimental data, even if the recirculation bubble is overly reduced in size. For the pressure field, improvements are observed even if they are not quantitatively important as for the velocity field. This may be due to the segregated structure of the CFD solver, which employs the pressure as a Lagrangian multiplier to impose the incompressibility constraint. Potentially, more sophisticated coupled solvers could provide improved results when used in DA tools using pressure data as observation.

Future investigations include more sophisticated parametric descriptions of the turbulence modelling employed, including coupling between DA tools and machine learning applications.

This work was granted access to the HPC resources of GENCI in the framework of the resources requested in A12 for project A0122A01741 on the IRENE supercomputer (TGCC). Florent Duchaine and Miguel Ángel Moratilla-Vega are warmly acknowledged for the help provided during the early stages of development of CONES. 






\bibliography{bibliography}


\end{document}